\documentclass[10pt,twocolumn,letterpaper]{article}

\usepackage{cvpr}              
\usepackage[accsupp]{axessibility}  

\usepackage{multirow}
\usepackage{bm}
\usepackage{overpic}
\usepackage{pict2e} 

\newcommand{\p}{\bm{\mathrm{p}}}
\newcommand{\x}{\bm{\mathrm{x}}}
\newcommand{\n}{\bm{n}}

\newcommand{\wh}{\bm{h}}
\newcommand{\whr}{\bm{h}_\mathrm{r}}
\newcommand{\wht}{\bm{h}_\mathrm{t}}
\newcommand{\wi}{\bm\omega_\mathrm{i}}

\newcommand{\wo}{\bm\omega_\mathrm{o}}

\newcommand{\dotp}[2]{|#1 \cdot #2|}
\newcommand{\f}{f}
\newcommand{\fb}{f_\mathrm{\beta}}
\newcommand{\fd}{f_\mathrm{d}}
\newcommand{\fs}{f_\mathrm{s}}
\newcommand{\ft}{f_\mathrm{t}}
\newcommand{\Fb}{M_\mathrm{\beta}}
\newcommand{\Fd}{M_\mathrm{d}}
\newcommand{\Fs}{M_\mathrm{s}}
\newcommand{\Ft}{M_\mathrm{t}}
\newcommand{\Lb}{E_\mathrm{\beta}}
\newcommand{\Ld}{E_\mathrm{d}}
\newcommand{\Ls}{E_\mathrm{s}}
\newcommand{\Lt}{E_\mathrm{t}}
\newcommand{\kd}{k_\mathrm{d}}
\newcommand{\ks}{k_\mathrm{s}}
\newcommand{\kt}{k_\mathrm{t}}

\newcommand{\Rbbb}{\mathbb{R}^{H\times W\times 3}}
\newcommand{\Rbb}{\mathbb{R}^{H\times W}}
\newcommand{\Sph}{{{\mathcal S}^2}}
\newcommand{\Hem}{\mathcal H^2_+}
\newcommand{\hem}{\mathcal H^2_-}
\newcommand{\intd}{\,\mathrm{d}}
\newcommand{\kernel}{\mathcal K}

\newcommand{\I}{\bm{\mathrm{I}}}
\newcommand{\Id}{\bm{\mathrm{I}}_\mathrm{diff}}
\newcommand{\Is}{\bm{\mathrm{I}}_\mathrm{spec}}
\newcommand{\It}{\bm{\mathrm{I}}_\mathrm{tran}}
\newcommand{\X}{\bm{\mathrm{X}}}
\newcommand{\D}{\bm{\mathrm{D}}}
\newcommand{\N}{\bm{\mathrm{N}}}
\newcommand{\A}{\bm{\mathrm{A}}}
\newcommand{\M}{\bm{\mathrm{M}}}
\newcommand{\R}{\bm{\mathrm{R}}}
\newcommand{\T}{\bm{\mathrm{T}}}
\newcommand{\E}{\bm{\mathrm{E}}}
\newcommand{\Amr}{\bm{\mathrm{A}}_\mathrm{mr}}
\newcommand{\Abg}{\bm{\mathrm{A}}_\mathrm{bg}}

\graphicspath{{fig/}}
\newlength{\resLen}
\newlength{\boxLen}

\newcommand\imwidth{}
\newcommand\imheight{}

\newcommand\xone{}
\newcommand\yone{}
\newcommand\wone{}
\newcommand\hone{}

\newcommand\xtwo{}
\newcommand\ytwo{}
\newcommand\wtwo{}
\newcommand\htwo{}

\newcommand\xthr{}
\newcommand\ythr{}
\newcommand\wthr{}
\newcommand\hthr{}

\newcommand\xfou{}
\newcommand\yfou{}
\newcommand\wfou{}
\newcommand\hfou{}

\newcommand\xfiv{}
\newcommand\yfiv{}
\newcommand\wfiv{}
\newcommand\hfiv{}

\newcommand\leftone{}
\newcommand\botone{}
\newcommand\rightone{}
\newcommand\topone{}
\newcommand\bxone{}
\newcommand\bxxone{}
\newcommand\byone{}
\newcommand\byyone{}

\newcommand\lefttwo{}
\newcommand\bottwo{}
\newcommand\righttwo{}
\newcommand\toptwo{}
\newcommand\bxtwo{}
\newcommand\bxxtwo{}
\newcommand\bytwo{}
\newcommand\byytwo{}

\newcommand\leftthr{}
\newcommand\botthr{}
\newcommand\rightthr{}
\newcommand\topthr{}
\newcommand\bxthr{}
\newcommand\bxxthr{}
\newcommand\bythr{}
\newcommand\byythr{}

\newcommand\leftfou{}
\newcommand\botfou{}
\newcommand\rightfou{}
\newcommand\topfou{}
\newcommand\bxfou{}
\newcommand\bxxfou{}
\newcommand\byfou{}
\newcommand\byyfou{}

\newcommand\leftfiv{}
\newcommand\botfiv{}
\newcommand\rightfiv{}
\newcommand\topfiv{}
\newcommand\bxfiv{}
\newcommand\bxxfiv{}
\newcommand\byfiv{}
\newcommand\byyfiv{}

\definecolor{cvprblue}{rgb}{0.21,0.49,0.74}
\usepackage[pagebackref,breaklinks,colorlinks,allcolors=cvprblue]{hyperref}

\def\paperID{15} 
\def\confName{CVPR Workshop}
\def\confYear{2025}

\title{ePBR: Extended PBR Materials in Image Synthesis}

\author{Yu Guo, Zhiqiang Lao, Xiyun Song, Yubin Zhou, Zongfang Lin and Heather Yu\\
Futurewei Technologies, US\\
{\tt\small \{yguo1,zlao,xsong,yzhou2,zlin1,hyu\}@futurewei.com}}

\begin{document}
\maketitle
\begin{abstract}
Realistic indoor or outdoor image synthesis is a core challenge in computer vision and graphics. The learning-based approach is easy to use but lacks physical consistency, while traditional Physically Based Rendering (PBR) offers high realism but is computationally expensive. Intrinsic image representation offers a well-balanced trade-off, decomposing images into fundamental components (intrinsic channels) such as geometry, materials, and illumination for controllable synthesis. However, existing PBR materials struggle with complex surface models, particularly high-specular and transparent surfaces. In this work, we extend intrinsic image representations to incorporate both reflection and transmission properties, enabling the synthesis of transparent materials such as glass and windows. We propose an explicit intrinsic compositing framework that provides deterministic, interpretable image synthesis. With the Extended PBR (ePBR) Materials, we can effectively edit the materials with precise controls.
\end{abstract}    
\section{Introduction}
\label{sec:intro}

Image synthesis is a fundamental challenge in computer vision and graphics, with applications ranging from content generation to realistic scene manipulation. Existing synthesis methods can be broadly classified into three main approaches, each offering different levels of control, realism, and computational efficiency.

Physically Based Rendering (PBR) generates images by simulating the physical interaction of light with scene geometry, materials, and illumination. This method provides highly realistic results by ensuring consistency with real-world physics. However, it requires a complete 3D scene representation and is computationally expensive, making it impractical for scenarios where only a single image is available or when real-time performance is needed.

Learning-based direct image generation leverages deep generative models such as diffusion models to synthesize images based on learned priors. These models can efficiently generate high-quality and diverse images, making them suitable for applications such as texture synthesis, artistic rendering, and creative content generation. However, since these approaches do not rely on explicit scene representations, they lack precise control over scene properties such as geometry, material consistency, and lighting, often leading to physically implausible results.

\begin{figure}[t]
    \centering
    \setlength{\resLen}{0.24\linewidth}
    \addtolength{\tabcolsep}{-5pt}
    \begin{tabular}{cccc}
        \includegraphics[width=\resLen]{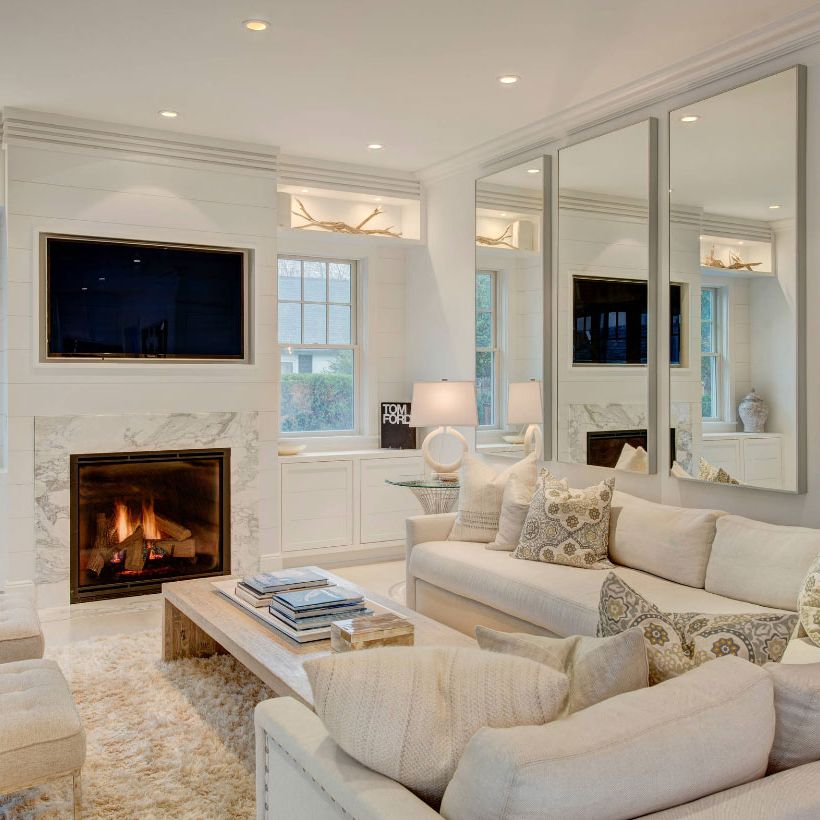} &
        \includegraphics[width=\resLen]{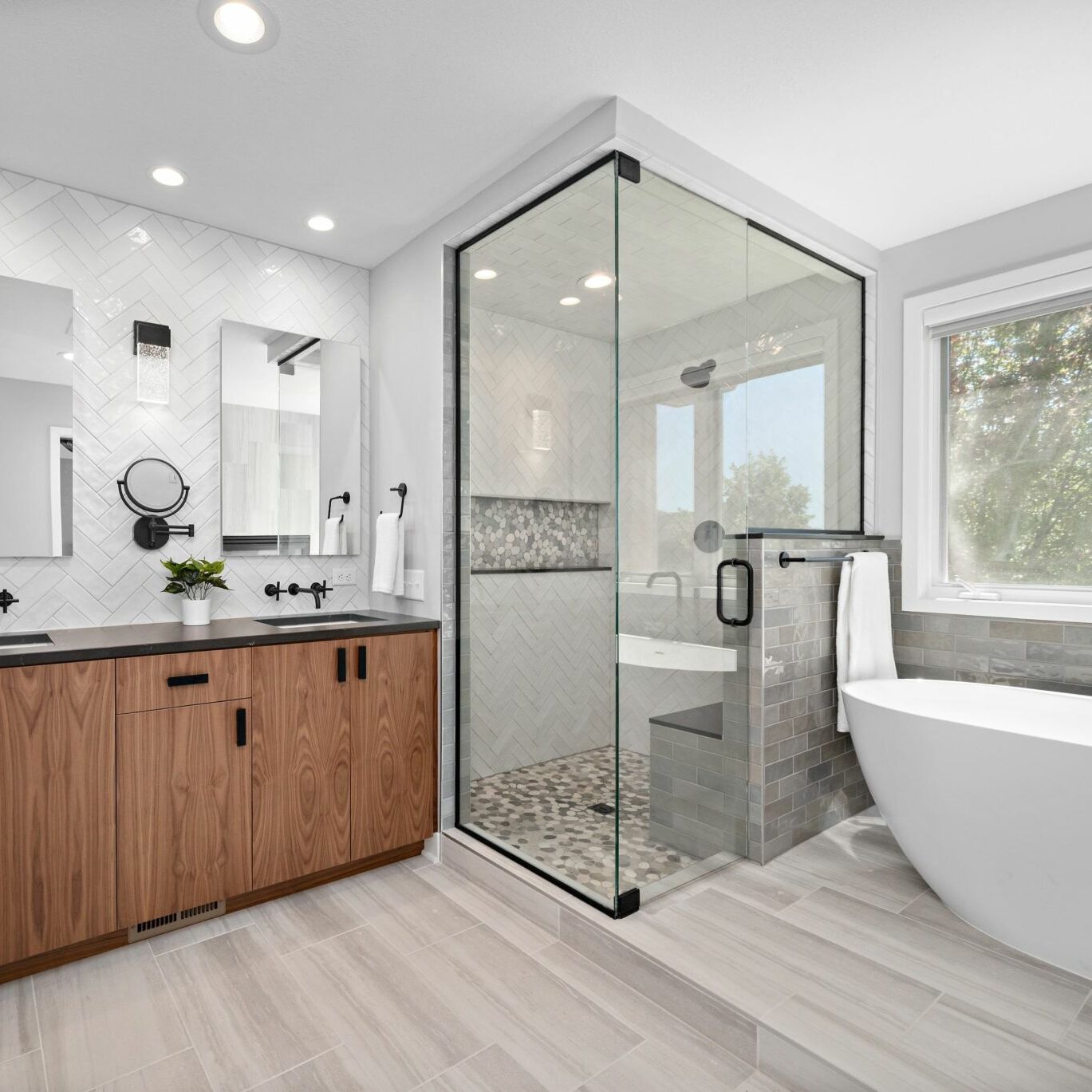} &
        \includegraphics[width=\resLen]{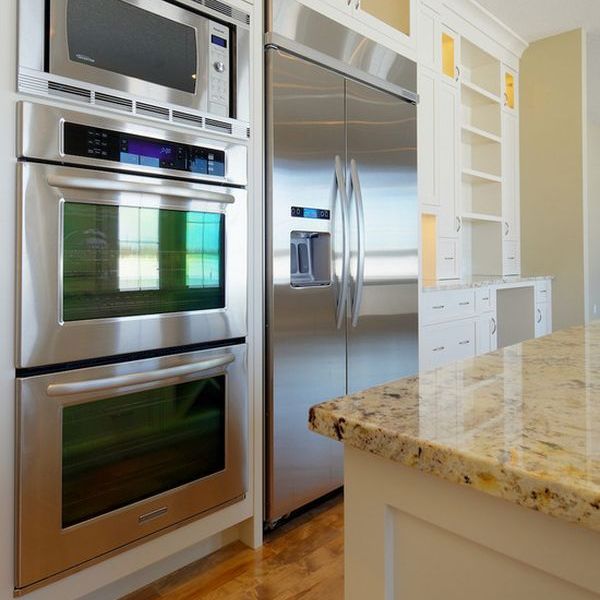} & 
        \includegraphics[width=\resLen]{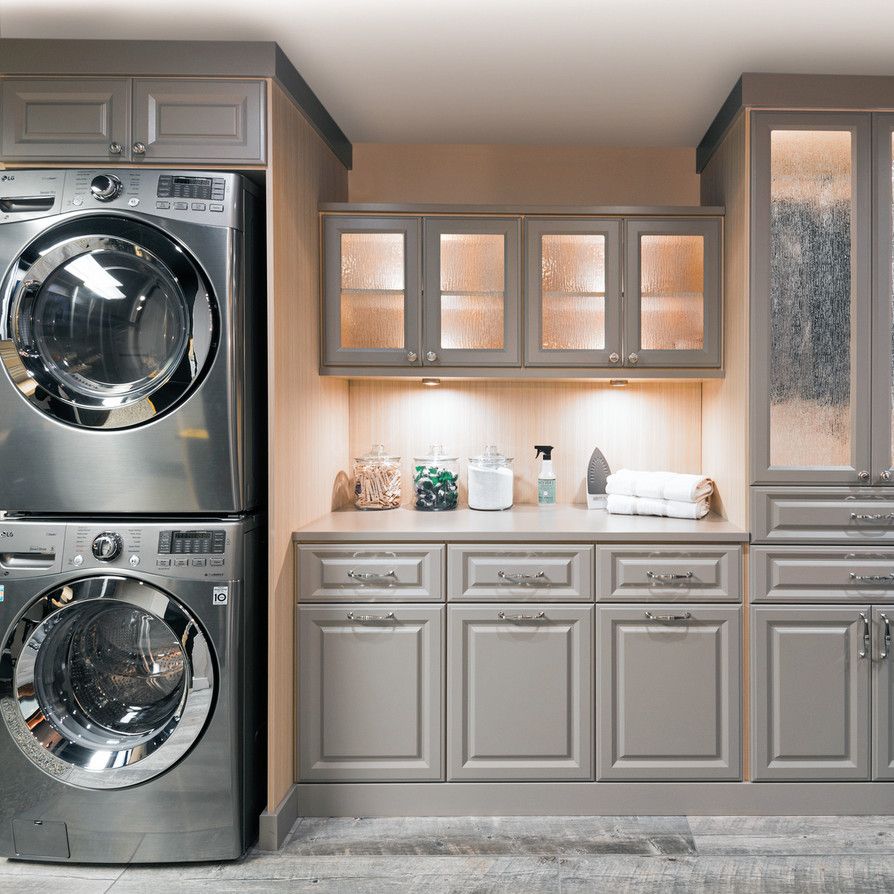} 
        \\[-1pt]
        \includegraphics[width=\resLen]{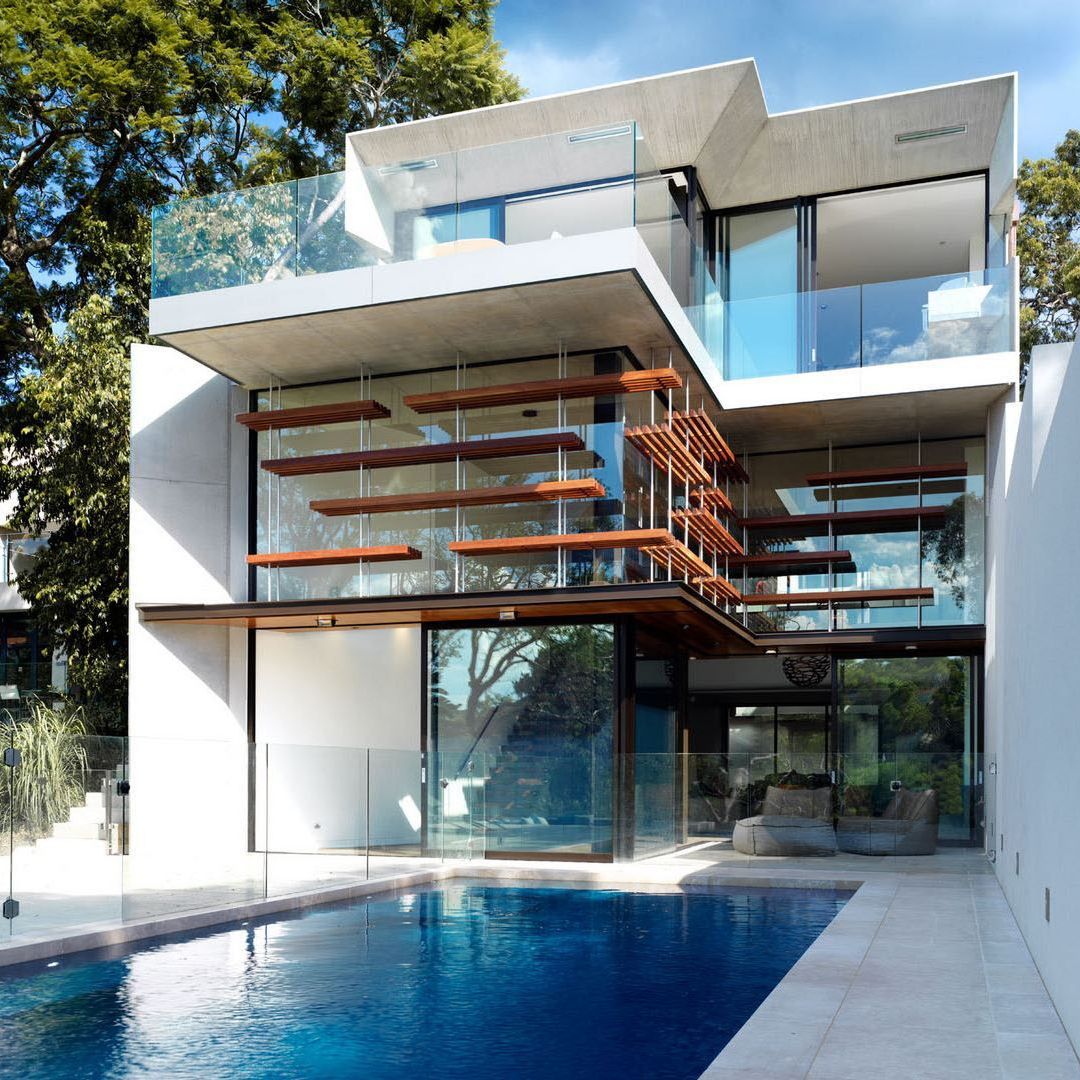} &
        \includegraphics[width=\resLen]{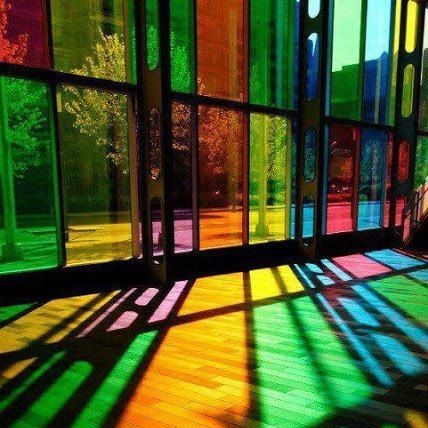} &
        \includegraphics[width=\resLen]{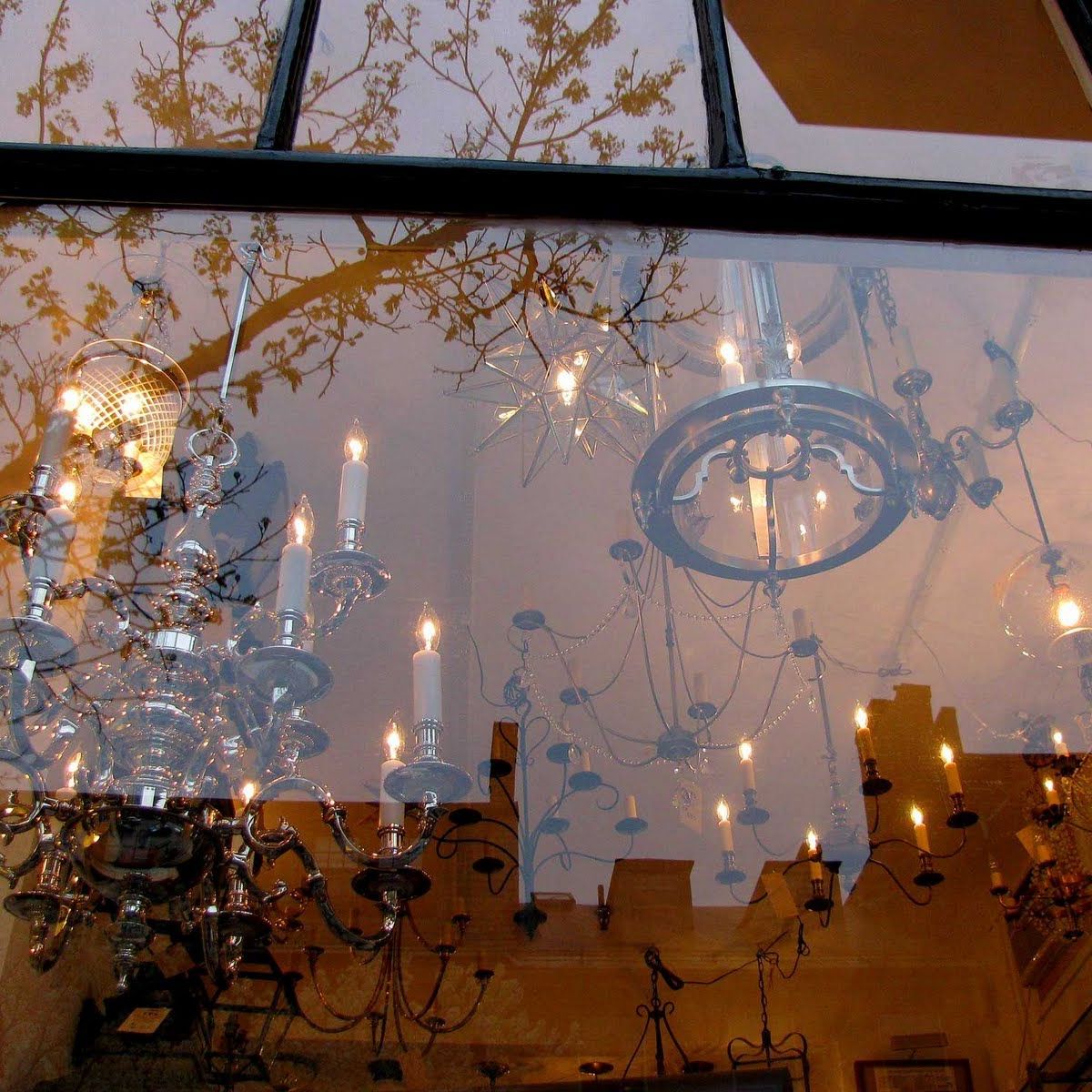} &
        \includegraphics[width=\resLen]{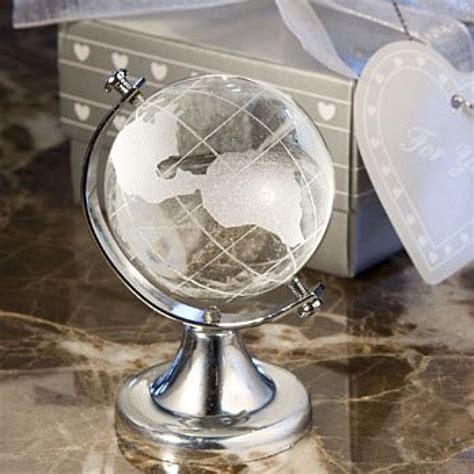} 
    \end{tabular}\vspace{-8pt}
    \caption{
        Highly specular and transparent objects are very common in the real world. 
    }\vspace{-8pt}
    \label{fig:gallery}
\end{figure}

Intrinsic representation for image synthesis offers a hybrid approach that balances realism and controllability. Composing an image into intrinsic channels, such as geometry, materials, shading, and illumination, provides a structured representation that enables flexible image manipulation while maintaining a meaningful connection to physical properties. Unlike full PBR rendering, which requires complex simulations, intrinsic representations can synthesize images efficiently by modifying specific scene attributes and recombining them. This approach is widely adopted in real-time rendering applications, such as video games, where intrinsic channels facilitate material and lighting adjustments without expensive re-rendering.

Despite advances in intrinsic decomposition, current models often struggle to handle complex materials commonly found in real-world environments, particularly in indoor scenes. Traditional reflection models, such as Lambertian reflectance for diffuse surfaces and microfacet models for glossy surfaces, fail to capture highly specular and transparent materials such as glass, mirrors, and polished metals (see Fig. \ref{fig:gallery}). We propose an extended intrinsic representation incorporating reflection and transmission properties to address these challenges, allowing for a more comprehensive material synthesis.

By leveraging these enhanced intrinsic channels, our approach enables more accurate and controllable image synthesis, bridging the gap between physically-based methods and AI-driven generative models. This framework provides a structured yet efficient way of synthesizing diverse and realistic images while maintaining control over scene properties. It is well suited for applications in virtual content creation, digital art, and photorealistic rendering. Our contributions can be summarized as

\begin{itemize}
    \item We extend the intrinsic representation of indoor images from reflection only to transmission, which could model transparent/translucent materials like windows or glass doors. 
    \item We provide an explicit intrinsic compositing method to render the image. Compared to \textit{diffusion}-based rendering, ours provides more accurate controls on high specular regions, and the output is deterministic. The composed image could be used as quick feedback since there is no limitation on GPU memory and image resolution, and only basic image operations are used here.   
\end{itemize}

This paper focuses on intrinsic channels in the screen space $\X$ (followed by RGB$\leftrightarrow$X \cite{zeng2024rgb}), and to verify the robustness of $\X$, we provide an analytic solution for image composition. Future work will explore replacing \textit{PBR materials} with our \textit{Extended PBR materials} in image decomposition.
\section{Related works}
\label{sec:related}

\subsection{Physically based rendering}
Physically Based Rendering (PBR) has been a foundational approach in computer graphics, focusing on simulating the interaction of light with complex scene descriptions consisting of geometry, materials, and illumination. Traditional rendering pipelines heavily rely on Monte Carlo (MC) light transport simulation \cite{veach1998robust, pharr2023physically}, which despite its accuracy, introduces stochastic noise due to limited sampling. 
Another key challenge in PBR is the representation of materials. A variety of real-world surfaces could be modeled by the Bidirectional Scattering Distribution Function (BSDF). One of the most influential solutions is the Disney Principled BSDF \cite{burley2012physically, burley2015extending}, which is largely based on physical principles and empirical observations. Although not strictly enforcing physical accuracy, the Disney BSDF provides a practical and intuitive parameterization of plausible materials, prioritizing artistic control and ease of use over strict physical correctness. This approach has significantly influenced material systems across commercial and open-source rendering engines, including Blender’s Principled BSDF \cite{blender}, Unreal Engine's Physically Based Materials \cite{ue5}, Mitsuba3's Principled and Thin Principled BSDF \cite{jakob2022mitsuba3}, et al. Its widespread adoption underscores its effectiveness in balancing realism and creative flexibility within PBR frameworks.
However, real-world materials are more than a simple surface model. Due to the complexity of the materials, researchers use specific appearance models to represent them, for example, layered objects \cite{jakob2014comprehensive, belcour2018efficient, guo2018position}, hair \cite{marschner2003light, chiang2015practical}, cloth \cite{montazeri2020practical, wang2022spongecake}, and iridescence material \cite{belcour2017practical, guillen2020general}.

\subsection{Data-driven image synthesis}
Recent advancements in image synthesis have explored alternative paradigms, mainly through generative models, which diverge from classical rendering methodologies. In particular, large-scale diffusion models have demonstrated remarkable success in producing highly realistic images by iteratively refining noise into structured visual content, e.g, DALL$\cdot$E \cite{ramesh2022hierarchical} and Stable Diffusion \cite{rombach2022high}. These models extend the neural denoising approach to its extreme, leveraging probabilistic image generation from pure Gaussian noise. Unlike traditional PBR, which requires explicit scene descriptions, \textit{diffusion}-based approaches learn complex visual distributions from extensive datasets, synthesizing diverse and photorealistic imagery without explicit physical simulation.
However, such models are hard to train and generated images are hard to control. Most of the other works try to fine-tune the pre-trained models for various domains and conditioning. 
ControlNet \cite{zhang2023adding} has been widely adopted for tasks requiring precise layout control, including sketch-to-image synthesis, depth-aware image generation, and pose-conditioned rendering. IC-Light \cite{zhang2025scaling} changes the illumination of the image but retains the underlying image details and maintains the intrinsic properties unchanged. 
More works that use diffusion models for appearance and illumination manipulation are summarized in the related works of IC-Light \cite{zhang2025scaling} and a survey \cite{huang2025diffusion}.

\subsection{Intrinsic representation}
While accurate rendering realistic images from explicit scene descriptions is often computationally expensive and labor intensive. In contrast, generating images from learned data through deep generative models offers efficiency but lacks precise control over the structure of the scene. Intrinsic representation provides a promising middle ground by balancing accuracy and flexibility, enabling structured yet editable scene decomposition. A well-designed intrinsic representation should capture fundamental physical properties such as geometry, materials, and illumination while remaining intuitive for manipulation.
In synthetic image generation, particularly in real-time rendering pipelines, intrinsic properties can be efficiently extracted from render buffers, making them readily available for downstream tasks such as relighting and material editing. These representations have also been widely integrated into modern neural rendering and inverse graphics frameworks, facilitating explicit control over scene components for high-quality synthesis and manipulation. By bridging physically-based rendering with data-driven approaches, intrinsic representations play a crucial role in achieving both photorealism and user-controllable scene generation.

Intrinsic geometry representations focus on recovering structural scene information, such as normal and depth maps. Traditional methods rely on photometric constraints, while neural approaches use large datasets to predict detailed geometric properties from single images \cite{yang2024depth, yang2025depth, ye2024stablenormal, he2024lotus}.

The intrinsic lighting representation aims to estimate scene illumination separately from geometry and reflectance. Various representations are used for different tasks, such as environment maps \cite{wang2025materialist, liang2025diffusionrenderer}, spherical harmonics \cite{ramamoorthi2001efficient}, and lighting fields \cite{li2020inverse, zhu2022learning}.

Reflectance-shading is one of the most studied forms of intrinsic representation. This approach separates an image into reflectance (albedo) and shading components \cite{careaga2023intrinsic, luo2024intrinsicdiffusion}. Reflectance captures the inherent color of surfaces, independently of lighting, while shading encodes the interaction of light with surface geometry. \cite{careaga2024colorful} further separates it to diffuse shading and specular residual. 

The most common representation in recent research is using the parametric microfacet Bidirectional Reflectance Distribution Function (BRDF) with PBR. In addition to albedo, intrinsic representations also include additional properties of the material such as roughness, metallicity, or specularity, which are also called \textit{PBR materials}. It is widely used in single object tasks, such as 3D reconstruction and texture generation \cite{zhang2024dreammat, huang2024material, wang2024boosting, siddiqui2025meta, zhu2024mcmat}. It is also used in estimating spatially varying BRDFs for complex materials but with metallicity replaced by specularity \cite{deschaintre2018single, guo2020materialgan, ma2023opensvbrdf, wang2024nfplight}. Now, the representation is expanded to decomposition and editing of indoor images \cite{zeng2024rgb, kocsis2024intrinsic, zhu2022learning, li2020inverse} and even for videos \cite{liang2025diffusionrenderer}.

Intrinsic representations play a key role in bridging physically based rendering and generative models, enabling interpretable, editable, and physically consistent scene decomposition. Their continued development contributes to advances in inverse rendering, neural rendering, and material estimation, expanding the capabilities of traditional and learning-based graphics pipelines.

\subsection{Transparent/translucent material}
Limited types of materials with only diffuse and specular reflectance are used in image manipulation. We want to extend this to transparent materials for both light reflection and transmission, which would take into consideration the Bidirectional Transmittance Distribution Function (BTDF). Only a few works consider glass-like materials. 
Materialist \cite{wang2025materialist} supports adding transparent objects in the image, but they only considered it as a special case. The Alchemist \cite{sharma2024alchemist} can control the transparency of a single object with a learning approach. \cite{ma2023opensvbrdf, rodriguez2025single} takes transparency as an intrinsic map in SVBRDF to better support fabrics.

\section{Method}
In this section, we will combine BRDF and BTDF with the thin-surface assumption to achieve a full BSDF model. Starting from rendering equation (\cref{sec:re}), we first go through BRDF model (\cref{sec:brdf}), then BTDF (\cref{sec:btdf}), and at the end we get our final BSDF model (\cref{sec:epbr}).  

\subsection{Rendering Equation}
\label{sec:re}

The rendering equation \cite{veach1998robust} for a non-emissive surface point $\p$ is,
\begin{equation}
    \label{eqn:rendering}
    L(\p, \wo) = \int_\Sph \f(\p, \wo, \wi) \, L(\p, \wi) \, \dotp{\wi}{\n_{\p}} \intd \wi
\end{equation}
where $L(\p, \wo)$ is the outgoing radiance at position $\p$ in the viewing direction of $\wo$ ($\p$ to camera origin). $L(\p, \wi)$ is the irradiance at $\p$ with the direction of the light leaving $\p$. Radiance at $\p$ is the integration of irradiance in all directions of a sphere with weights of $\f(\p, \wo, \wi)$, which is the BSDF at $\p$. And $\n_{\p}$ is the surface normal of $\p$. \footnote{$\p$ will be dropped for simplification.}

Disney Principled BSDF \cite{burley2012physically, burley2015extending} is the widely used appearance model in all kinds of rendering engines since it can support many types of materials and lighting effects, including diffuse, subsurface scattering, retroreflective, specular reflectance, clear coating, transmissive surface with refraction, and sheen compensation for retroreflective.

The BRDF used in recent research works is a simplified Disney Principled BSDF model, which can only represent a surface's diffuse and specular reflection.
This section will extend it to a thin-surface model that can handle both reflection and transmission.
\begin{equation}
    \f = \kd\fd + \ks\fs + \kt\ft
\end{equation}
where $\kd$, $\ks$ and $\kt$ are the coefficients of diffuse ($\fd$), specular reflectance ($\fs$) and specular transmittance ($\ft$) terms.

\subsection{Reflection only surface}
\label{sec:brdf}

We can use the Lambertian model \cite{oren1994generalization} to estimate the diffuse reflectance ($a$ is the albedo of the surface) without considering subsurface scattering and grazing retroreflective,
\begin{equation}\label{eqn:diff}
    \fd = \frac{a}{\pi}    
\end{equation}

We ignore the clearcoat and only use one microfacet model \cite{cook1982reflectance} for specular reflectance,  
\begin{equation}\label{eqn:spec}
    \fs = \frac{D(\whr) \, F(\whr, \wo) \, G(\whr, \wo, \wi)}{4 \, \dotp{\wo}{\n} \, \dotp{\wi}{\n}}   
\end{equation}
$\whr$ is the half vector between $\wo$ and $\wi$, which is $\whr = (\wo + \wi)/(||\wo + \wi||)$. 

\noindent\textbf{Normal distribution function ($D$)}, also known as the specular distribution, describes the normal distribution of microfacets for the surface. 
We use GGX \cite{walter2007microfacet} distribution which is defined as follow ($r$ is the surface roughness),
\begin{equation}
    D(\whr) = \frac{r^4}{\pi \, (\dotp{\n}{\whr}^2 \, (r^4-1) + 1)^2}
\end{equation}

\noindent\textbf{Fresnel reflection coefficient ($F$)} describes the amount of light that reflects from a mirror surface given its Index of Refraction (IOR). 
For the Fresnel term, we use Schlick's approximation \cite{schlick1994inexpensive} ($F_0 = (1-\eta)^2/(1+\eta)^2$),
\begin{equation}
    F(\whr, \wo) = F_0 + (1 - F_0) \, (1 - \dotp{\wo}{\whr})^5
\end{equation}

\noindent\textbf{Geometric attenuation ($G$)} describes the shadowing from the microfacets.
We use Smith's method (independent of $\wh$) with Schlick approximation \cite{schlick1994inexpensive} for it ($k = r^2/2$),
\begin{equation}
    G(\whr, \wo, \wi) = \frac{\dotp{\n}{\wo}}{\dotp{\n}{\wo}\,(1-k)+k} \, \frac{\dotp{\n}{\wi}}{\dotp{\n}{\wi}\,(1-k)+k}
\end{equation}

\subsection{Transparent thin surface}
\label{sec:btdf}

For a general transparent object, like a bottle of orange juice, light refracts into the liquid and out from the other side after absorption and scattering. This process relies on the properties of two surfaces and the volumes between.

In this work, we focus only on thin surfaces with two parallel surfaces and zero thickness, which approximate an actual transparent surface, such as windows or a glass table.

\begin{figure}[t]
    \centering
    \setlength{\resLen}{0.7\linewidth}
    \includegraphics[width=\resLen]{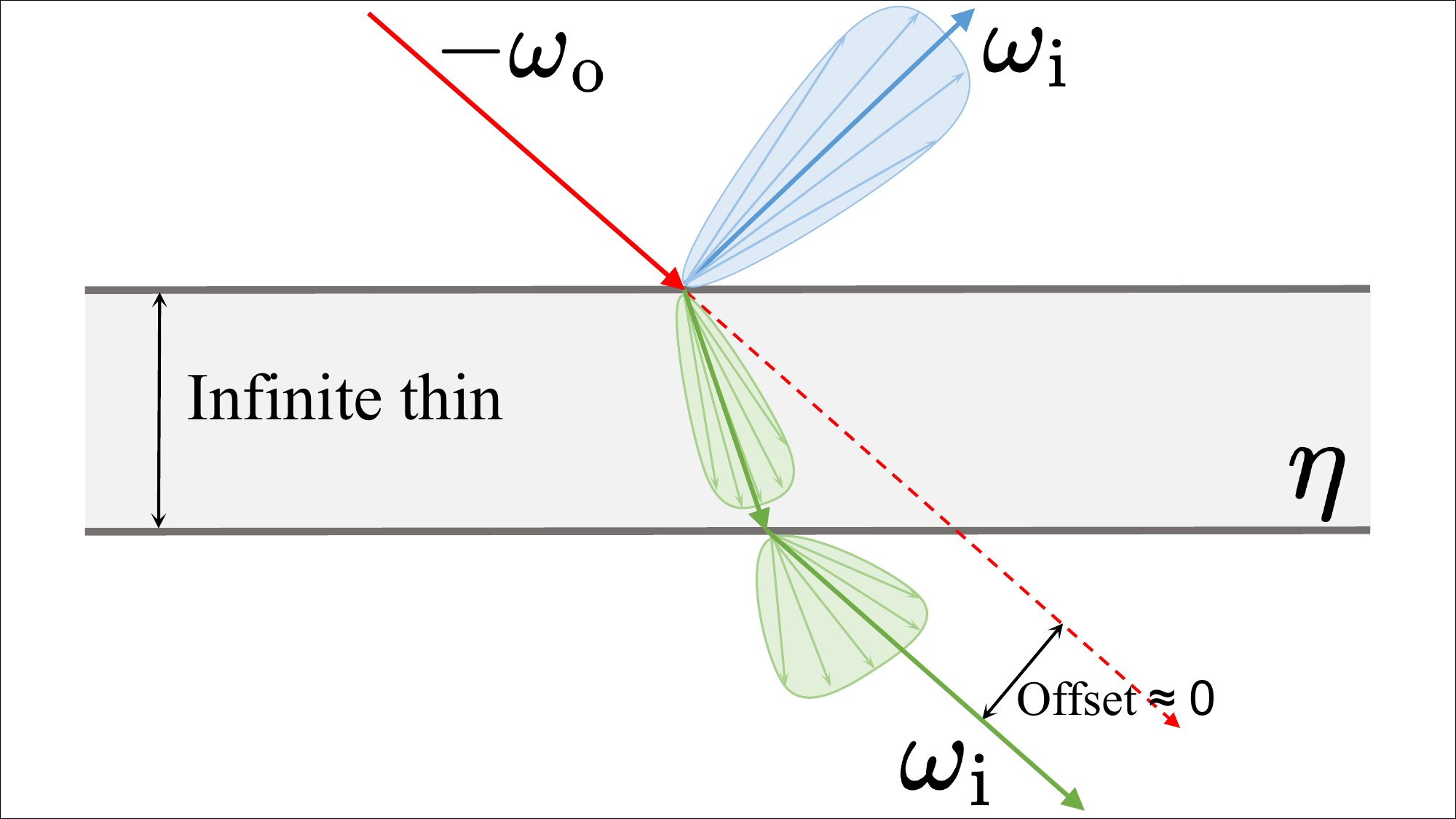}\vspace{-8pt}
    \caption{
        \textbf{Thin surface assumption:} Ignoring the internal reflection, light traveling through a transparent thin surface refracts twice as it enters and exits, and reflects once only on the top surface. For a smooth surface, light exits with the same direction as it enters and the offset could be ignored. 
    }\vspace{-8pt}
    \label{fig:thin}
\end{figure}

With the assumption of a thin surface (\cref{fig:thin}), we observe that light bending due to refraction approximately cancels, and the offset of incoming and outgoing light could be ignored. The specular transmission could be modeled by the microfacet distribution, the same as the specular lobe ($\fs$), but reflected to the other side. If we don't model the internal reflections between two surfaces while only considering two roughnesses, the transmission lobe could be written as,
\begin{equation}\label{eqn:trans}
    \ft = \frac{\hat{D}(\wht) \, F(\wht, \wo) \, G(\wht, \wo, \wi)}{4 \, \dotp{\wo}{\n} \, \dotp{\wi}{\n}}   
\end{equation}
$\hat{D}$ is the Extended Normal Distribution Function (eNDF), and can be estimated by joint spherical warping strategy \cite{guo2016rendering}.
$\wht$ is the half vector between $\wo$ and $\wi$, which is $\wht = -(\wo + \eta\wi)/(||\wo + \eta\wi||)$. 

\subsection{The ePBR material model}  
\label{sec:epbr}

We combine the three terms \cref{eqn:diff}, \cref{eqn:spec}, and \cref{eqn:trans} and weight them to get the final BSDF,
\begin{equation}
    \label{eqn:bsdf}
    \f = (1-t)(1-m)\fd + \fs + t\ft
\end{equation}
where $m$ is metallic and $t$ is transparency.
Albedo ($a$) is shared with the incident specular response to support metallic materials. So the $F_0$ is modified accordingly, $F_0 = \mathrm{lerp}(F_0, a, m)$. 

\begin{figure}[t]
    \centering
    \setlength{\resLen}{0.33\linewidth}
    \addtolength{\tabcolsep}{-5pt}
    \begin{tabular}{ccc}
        \includegraphics[width=\resLen]{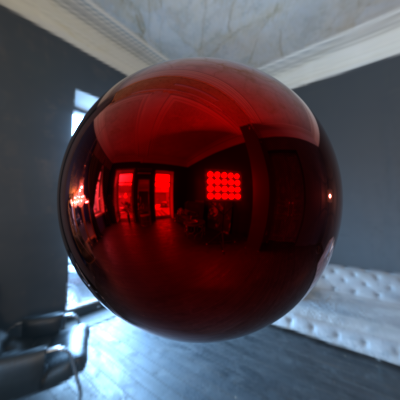} &
        \includegraphics[width=\resLen]{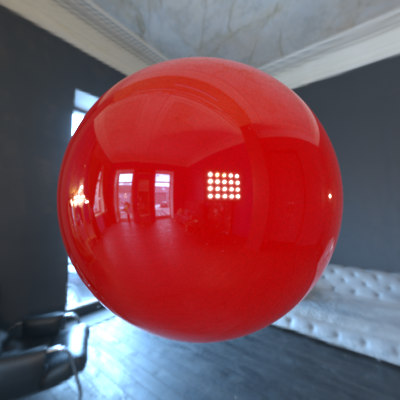} &
        \includegraphics[width=\resLen]{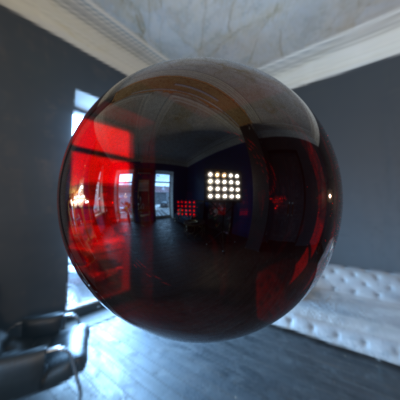} 
        \\
        \includegraphics[width=\resLen]{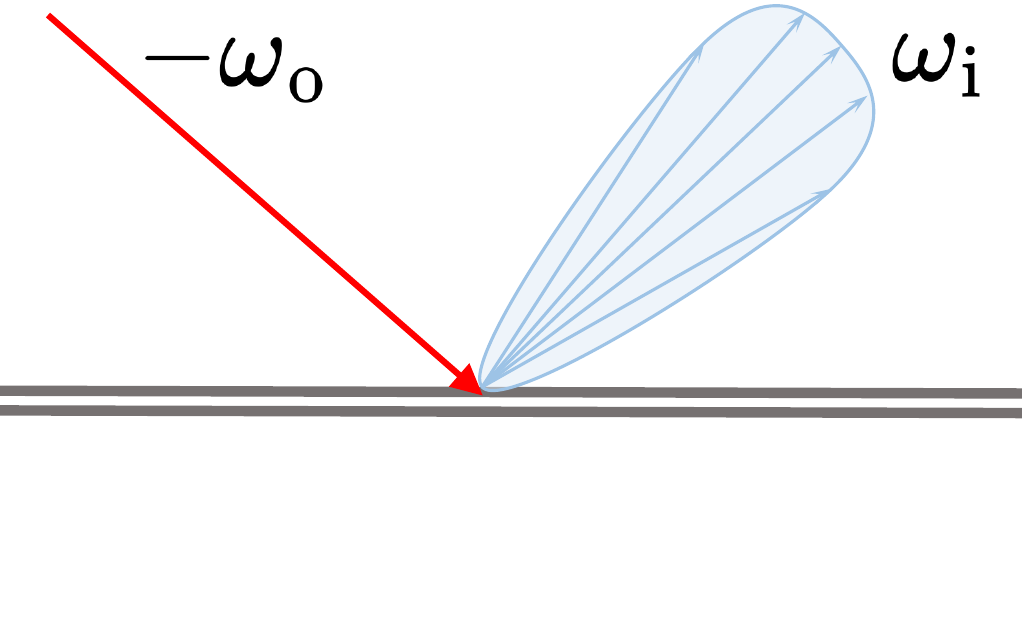} &
        \includegraphics[width=\resLen]{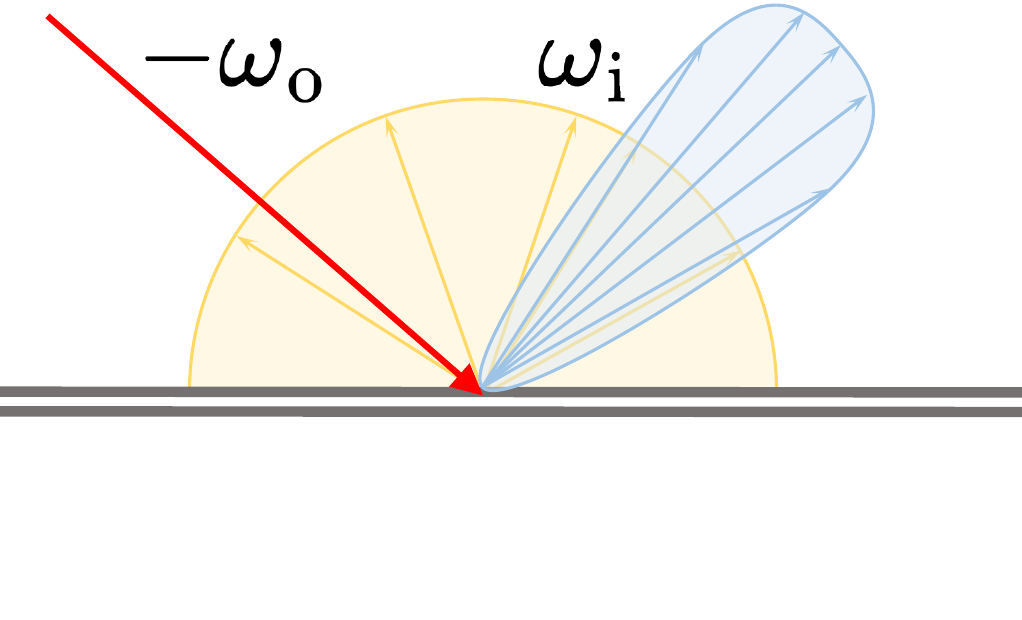} &
        \includegraphics[width=\resLen]{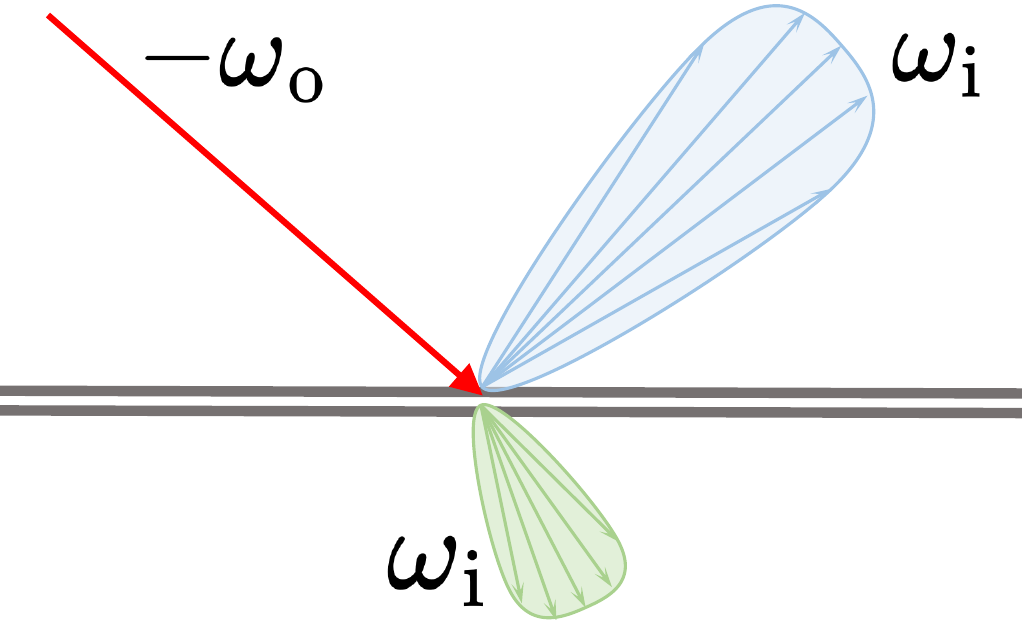} 
        \\
        Metal & Dielectric & Glass
    \end{tabular}\vspace{-8pt}
    \caption{
        \textbf{Three typical materials:} Metal ($t=0, m=1$), Dielectric ($t=0, m=0$) and Glass ($t=1, m=0$). 
    }\vspace{-8pt}
    \label{fig:key_type}
\end{figure}

\cref{fig:key_type} presents some special cases, and the linear mixture of them can model all other materials. \begin{itemize}
\item when $t=0$ and $m=0$, $\f = \fd + \fs(F_0=0.04)$, which indicates most dielectric materials in real life;
\item when $t=0$ and $m=1$, $\f = \fs(F_0=a)$, which is conductor/metal;
\item when $t=1$ and $m=0$, $\f = \fs(F_0=0.04) + \ft(F_0=0.04)$, which represents transparent glass;
\item when $t=1$ and $m=1$, transparent metal, which could be simulated using our model although it does not exist in nature. So, during image composition, we automatically set $m=0$ if $t>0$ to avoid getting invalid materials.
\end{itemize}

\section{Screen-space image synthesis}
In this section, we will explore how to use the intrinsic channels of the screen space $\X$ to synthesize the final image $\I$.  

First we put \cref{eqn:bsdf} back to \cref{eqn:rendering} and split it into three individual components,

\begin{equation}\begin{aligned}
    \label{eqn:rendering_bsdf}
    L(\wo) & = \kd\int_{\Hem} \fd(\wo, \wi) \, L(\wi) \, \dotp{\wi}{\n} \intd \wi \\
    & + \ks\int_{\Hem} \fs(\wo, \wi) \, L(\wi) \, \dotp{\wi}{\n} \intd \wi \\
    & + \kt\int_{\hem} \ft(\wo, \wi) \, L(\wi) \, \dotp{\wi}{\n} \intd \wi
\end{aligned}\end{equation}
where $\kd=(1-t)(1-m)$, $\ks=1$, $\kt=t$, and $\p$ is dropped for simplification. Energy conservation is not considered.

\begin{figure}[t]
    \centering
    \setlength{\resLen}{0.24\linewidth}
    \addtolength{\tabcolsep}{-5pt}
    \begin{tabular}{cc@{\hskip 4pt}cc}
        \multicolumn{2}{c}{Low roughenss} & \multicolumn{2}{c}{High roughness}
        \\
        \includegraphics[width=\resLen]{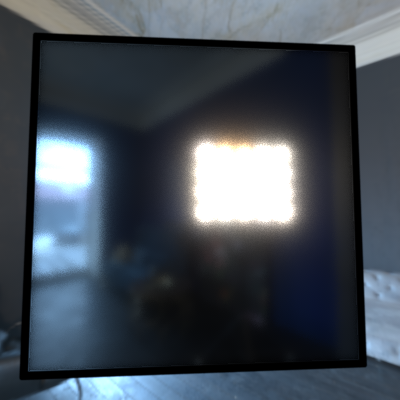} &
        \includegraphics[width=\resLen]{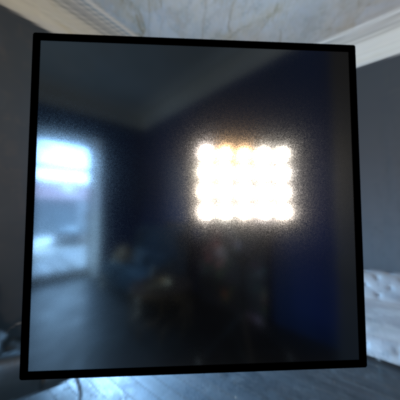} &
        \includegraphics[width=\resLen]{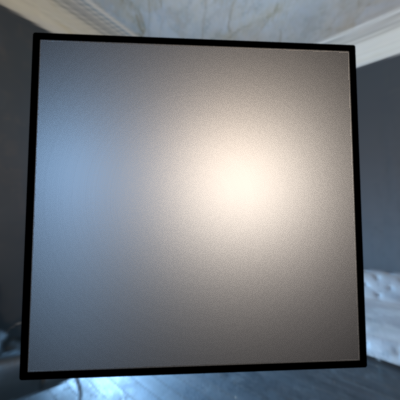} &
        \includegraphics[width=\resLen]{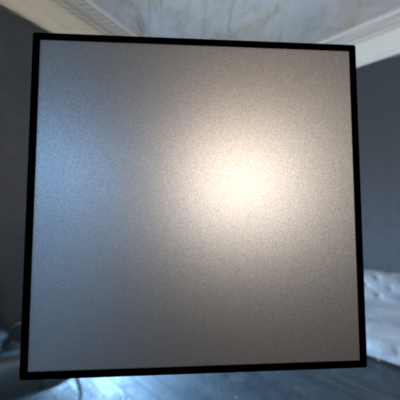} 
        \\
        Ours & GT & Ours & GT 
    \end{tabular}\vspace{-8pt}
    \caption{
        \textbf{Roughness evaluation.} Ours: Directly apply filtering kernel to the specular reflection image; GT: Path tracing with Monte Carlo sampling. 
    }\vspace{-18pt}
    \label{fig:rough}
\end{figure}

There is no analytical solution for the radiance integrals. Monte Carlo simulation with importance sampling is widely used to compute the color of a shading point $\p$ seen from a pixel $\x$. 

Inspired by Unreal Engine's Split-Sum method \cite{karis2013real}, and followed by \cite{zhuang2021real}, each radiance integral in \cref{eqn:rendering_bsdf} can be demodulated into two independent integrals, \textit{material} ($\Fb$) and \textit{weighted-lighting} ($\Lb$). 
\begin{equation}
    \Fb(\wo) =  \int \fb(\wo, \wi) \, \dotp{\wi}{\n} \intd \wi
\end{equation}
\vspace{-15pt}
\begin{equation}\begin{aligned}
    \Lb(\wo) & = \int W(\wo, \wi) \, L(\wi) \intd \wi \\
             & = \int \frac{\fb(\wo, \wi) \, \dotp{\wi}{\n}}{\int \fb(\wo, \wi) \, \dotp{\wi}{\n} \intd \wi} \, L(\wi) \intd \wi
\end{aligned}\end{equation}
where $\beta \in (\mathrm{b}, \mathrm{s}, \mathrm{t})$ is short for \textit{diffuse reflectance}, \textit{specular reflectance} and \textit{specular transmittance}.

\subsection{Diffuse reflectance}
For diffuse component, $\Fd$ can be calculated directly, 
\begin{equation}
    \Fd = \int \frac{a}{\pi} \, \dotp{\wi}{\n} \intd \wi = a   
\end{equation}
And its corresponding $\Ld$ is actually the diffuse irradiance,
\begin{equation}
    \Ld = \int \frac{\dotp{\wi}{\n}}{\pi} \, L(\wi) \intd \wi
\end{equation}
which is a commonly used G-buffer from deferred rendering \cite{pharr2005gpu}. It represents the amount of light reaching a shading point integrated over the upper cosine-weighted hemisphere and can be directly estimated from the input image \cite{zeng2024rgb}. 

So the diffuse reflectance $\Id$ is written as,
\begin{equation}
    \Id = \A \E
\end{equation}
note that both $a$ and $\A$ are albedo. The difference is that $a$ indicates the albedo of a single 3D surface point, while $\A$ is the albedo map in the screen space. $\E$ is the diffuse irradiance map. Without further notice, all the notation in capital letters and bold imply screen-space images. See \cref{sec:x} for a detailed definition.

\begin{figure}[t]
    \centering
    \setlength{\resLen}{0.9\linewidth}
    \begin{tabular}{c}
        \includegraphics[width=\resLen]{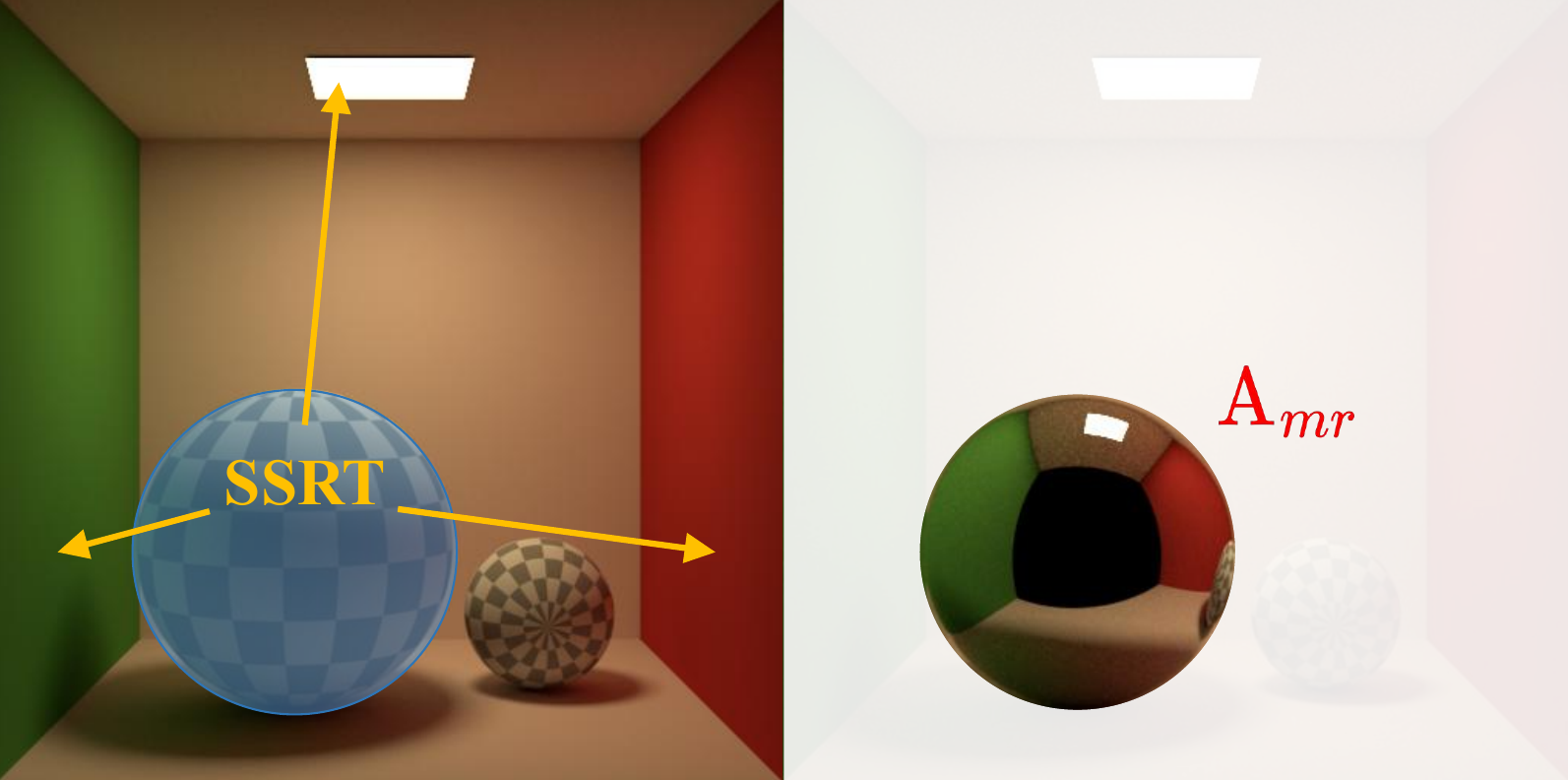}
    \end{tabular}\vspace{-8pt}
    \caption{
        \textbf{Generate $\Amr$ from SSRT.} To get a mirror-like reflection image for an interest region, using SSRT to trace the ray and find the corresponding color. 
    }\vspace{-8pt}
    \label{fig:ssrt}
\end{figure}

\subsection{Specular reflectance}
For specular component, we can convert $\Fs$ into a linear function of $F_0$ \cite{karis2013real},
\begin{equation}
    \Fs = \int \fs(\wo, \wi) \, \dotp{\wi}{\n} \intd \wi = A F_0 + B 
\end{equation}
The values of $A$ and $B$ depend on roughness $\R$. They could be precomputed and saved in a lookup table.

The corresponding $\Ls$ is the integral of the incident light weighted by the value $\fs\dotp{\wi}{\n}$. Instead of using importance sampling to convolve the environment map with the GGX distribution, we define a normalized filtering kernel $\kernel(\R,d)$ based on $D(\wh)\dotp{\wi}{\n}$. 
It has been verified that terms other than $D$ have relatively little effect on the shape of BSDF \cite{walter2007microfacet}. 
The shape of the kernel varies with roughness and shading distance (for simplicity, we use a fixed value for $d$ ), which can create blurring effects (see \cref{fig:rough}) when applying it to the mirror-like reflection image ($\Amr$). 
\begin{equation}
    \Ls = \textsc{Conv}(\kernel, \Amr)
\end{equation}

To obtain the reflection layer $\Amr$, we use Screen Space Ray Tracer (SSRT) \cite{zhu2022learning} to find the reflection color for each pixel $\x$. We take a depth map $\D$ and first compute the viewing direction for each point $\p$. With the normal map $\N$, it is easy to get the reflection ray. We trace the ray in the screen space to find the corresponding color and distance (see \cref{fig:ssrt}). 
\begin{equation}
    \Amr = \textsc{SSRT}(\D, \N)
\end{equation}

The $\Amr$ may have holes since the reflection ray may hit the back-face of an object in the scene or leave the screen space. Inpainting \cite{phongthawee2024diffusionlight} could be applied here if needed.

At the end, the specular reflectance $\Is$ can be written as,
\begin{equation}
    \Is =  (A F_0 + B) \; \textsc{Conv}(\kernel, \Amr)
\end{equation}

\subsection{Specular transmittance}
The way to compute transmittance ($\Ft$) is similar to $\Fs$. 
The main difference is that the kernel $\hat{\kernel}(\R,d) = \hat{D}(\wh)\dotp{\wi}{\n}$ that we applied to the background image ($\Abg$) has a wider distribution (\cite{dai2009dual, guo2016rendering}) since the light will go through both the top and bottom surfaces. In general, we assume the two surfaces have the same roughness, so instead of computing the accurate $\hat{\kernel}$, we simply apply it twice,
\begin{equation}
    \Lt = \textsc{Conv}(\kernel, \textsc{Conv}(\kernel, \Abg))
\end{equation}

If we want to add a transparent surface in a scene, we can easily get $\Abg$, but if we are going to change an opaque surface to transparent, image inpainting \cite{rombach2022high} is also needed here to guess what is behind the surface.

Now we have,
\begin{equation}
    \It =  (A F_0 + B)\cdot\textsc{Conv}(\kernel, \textsc{Conv}(\kernel, \Abg))\cdot\A
\end{equation}
notes that the surface albedo ($\A$) is multiplied here to approximate the absorption after light goes through the thin surface.

Finally, we combine the three layers to get the final image.
\begin{equation}
    \I = (1-\T)(1-\M)\Id + \Is + \T\It
\end{equation}

\section{Results}

\subsection{Intrinsic representation for ePBR materials}\label{sec:x}
This section summarizes the intrinsic channels $\X$ used in our model. (See \cref{tab:X} and \cref{fig:compose1} for example)

\newlength{\Oldarrayrulewidth}
\newcommand{\Cline}[2]{%
  \noalign{\global\setlength{\Oldarrayrulewidth}{\arrayrulewidth}}%
  \noalign{\global\setlength{\arrayrulewidth}{#1}}\cline{#2}%
  \noalign{\global\setlength{\arrayrulewidth}{\Oldarrayrulewidth}}}
\begin{table}[t]
    \centering
    \renewcommand{\arraystretch}{1.3}
    \hskip -7pt
    \begin{tabular}{c@{\hskip 5pt}c@{\hskip 8pt}l@{\hskip 8pt}c@{\hskip 8pt}l@{\hskip 1pt}c@{\hskip 1pt}c}
        \Cline{1pt}{2-5}
        & $\X$ & \multicolumn{1}{c}{Type} & Range & Description &
        \multirow{10}{*}[2ex]{$\left\}\rule{0pt}{5ex}\right.$ \hspace{-5pt} \rotatebox[origin=c]{90}{\footnotesize{PBR}}} &
        \multirow{10}{*}{$\left\}\rule{0pt}{7.5ex}\right.$ \hspace{-5pt} \rotatebox[origin=c]{90}{\footnotesize{ePBR}}}
        \\
        \cline{2-5}
        \multirow{2}{*}{\rotatebox[origin=c]{90}{\footnotesize{Geometry}}} & $\N$ & $\Rbbb$ & $[-1,1]$ & Normal \\ 
        & $\D$ & $\Rbb$ & $[0,1]$ & Depth \\ 
        \cline{2-5}
        \multirow{4}{*}{\rotatebox[origin=c]{90}{\footnotesize{Materials}}} & $\A$ & $\Rbbb$ & $[0,1]$ & Albedo \\ 
        & $\R$ & $\Rbb$ & $[0,1]$ & Roughness \\ 
        & $\M$ & $\Rbb$ & $[0,1]$ & Metallic \\ 
        & $\T$ & $\Rbb$ & $[0,1]$ & Transparency \\ 
        \cline{2-5}
        \multirow{3}{*}{\rotatebox[origin=c]{90}{\footnotesize{Illumination}}} & $\E$ & $\Rbbb$ & $[0,\infty)$ & Diffuse irradiance \\ 
        & $\Amr$ & $\Rbbb$ & $[0,\infty)$ & Mirror reflectance \\ 
        & $\Abg$ & $\Rbbb$ & $[0,\infty)$ & \small{Background radiance} \\ 
        \Cline{1pt}{2-5}
    \end{tabular}
    \caption{
        \textbf{The intrinsic channels} used in our model. Geometry: $\N$ and $\D$; Materials: $\A$ and $\R\M\T$; Illumination: $\E$, $\Amr$ and $\Abg$ (optional). 
    }
    \label{tab:X}
\end{table}


To be chosen as an intrinsic channel, it should follow some principles. The $\X$ must have the exact resolution of the image $\I$; each $\X$ should have its unique physical meaning, making precise image editing possible; the value of $\X$ should be uniformly distributed. Here is a bad example: If an intrinsic value ranges in $[0,1]$, the appearance changes dramatically when the value changes from $0$ to $0.1$, but there is almost no change if the value is greater than $0.1$. 

It always uses a 3-channel image to save PBR material \cite{zhu2022learning}. The \textit{red} channel refers to roughness ($\R$) and \textit{green} channel refers to metallic ($\M$), while \textit{blue} channel is always 0. Our ePBR model stores the transparency map ($\T$) in the \textit{blue} channel without additional memory cost.

\begin{figure*}[t]
    \centering
    \setlength{\resLen}{0.13\linewidth}
    \addtolength{\tabcolsep}{-5pt}
    \begin{tabular}{ccccccc}
        & \multicolumn{6}{c}{\hspace{3em} 0 \leftarrowfill \M \rightarrowfill 1 \;\;\;\;\;\;\;\;\;\;\;}
        \\
        \raisebox{30pt}{\rotatebox[origin=c]{90}{(a) Metallic}} &
        \includegraphics[width=\resLen]{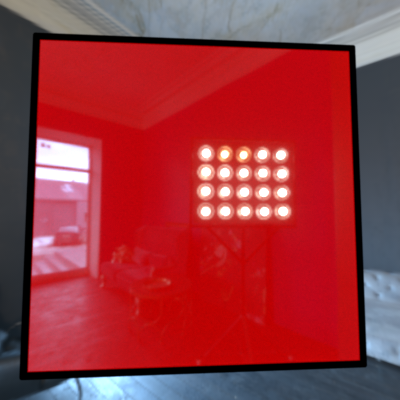} &
        \includegraphics[width=\resLen]{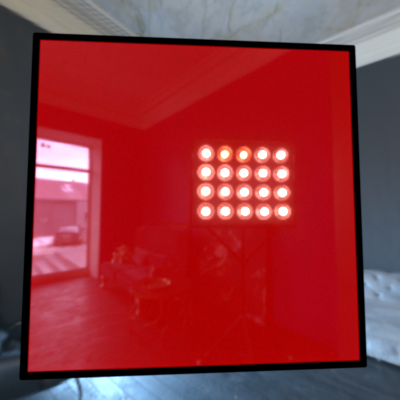} &
        \includegraphics[width=\resLen]{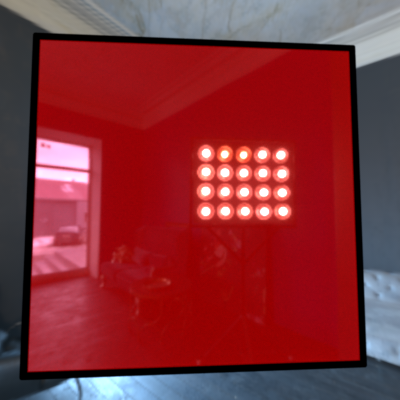} &
        \includegraphics[width=\resLen]{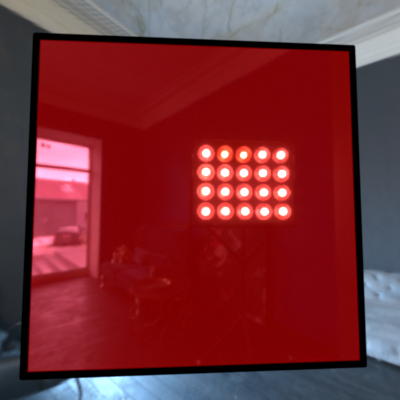} &
        \includegraphics[width=\resLen]{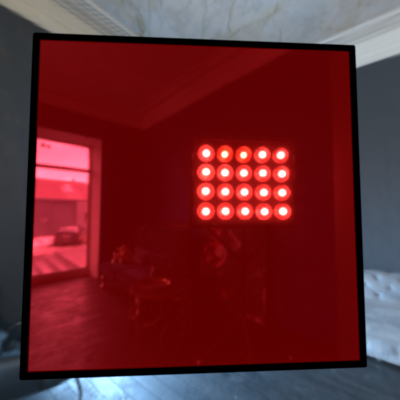} &
        \includegraphics[width=\resLen]{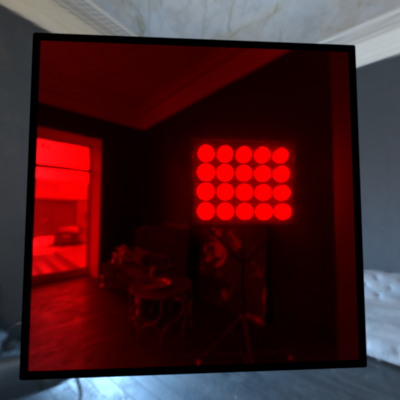} 
        \\
        & \multicolumn{6}{c}{\hspace{3em} 0 \leftarrowfill \R \rightarrowfill 0.5 \;\;\;\;\;\;\;\;\;}
        \\
        \raisebox{32pt}{\rotatebox[origin=c]{90}{(b) Roughness}} &
        \includegraphics[width=\resLen]{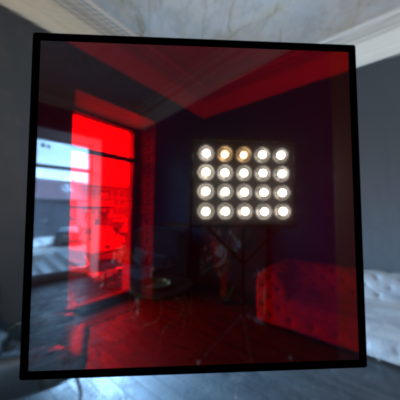} &
        \includegraphics[width=\resLen]{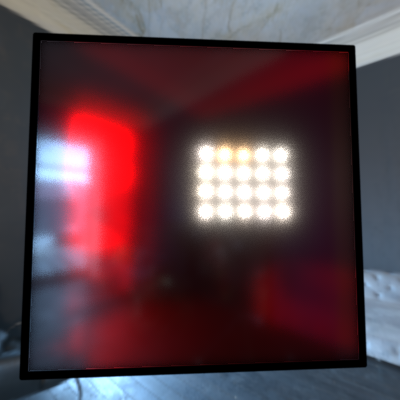} &
        \includegraphics[width=\resLen]{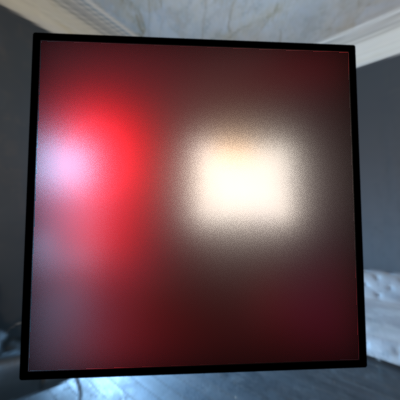} &
        \includegraphics[width=\resLen]{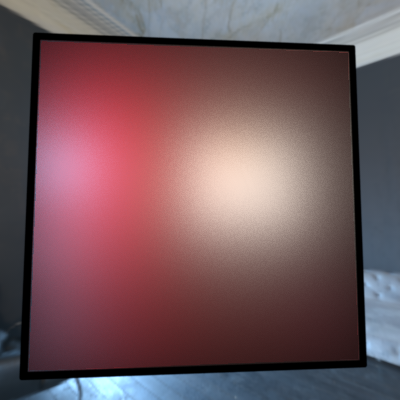} &
        \includegraphics[width=\resLen]{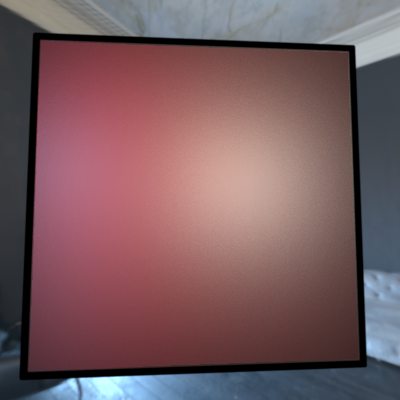} &
        \includegraphics[width=\resLen]{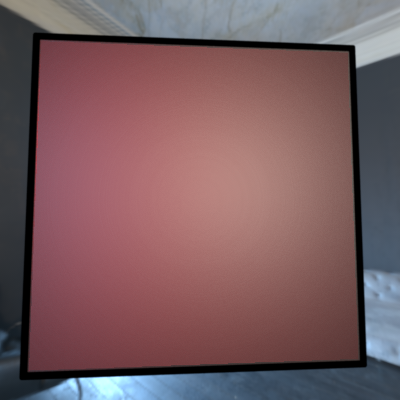} 
        \\
        & \multicolumn{6}{c}{\hspace{3em} 0 \leftarrowfill \T \rightarrowfill 1 \;\;\;\;\;\;\;\;\;\;\;}
        \\
        \raisebox{30pt}{\rotatebox[origin=c]{90}{(c) \small Transparency}} &
        \includegraphics[width=\resLen]{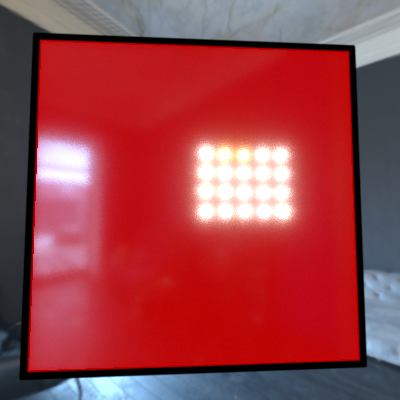} &
        \includegraphics[width=\resLen]{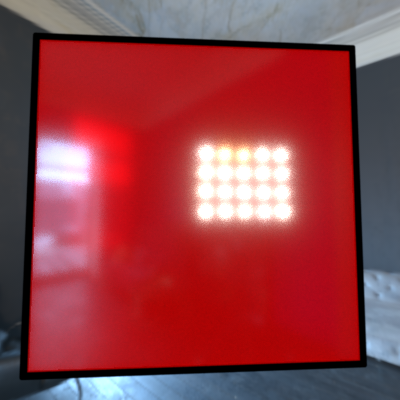} &
        \includegraphics[width=\resLen]{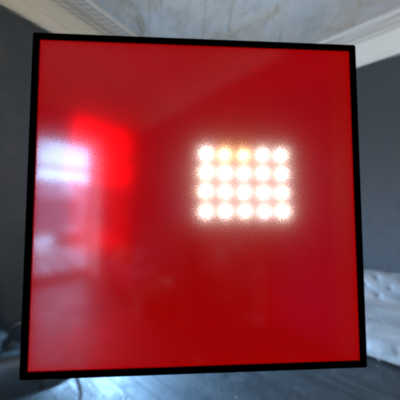} &
        \includegraphics[width=\resLen]{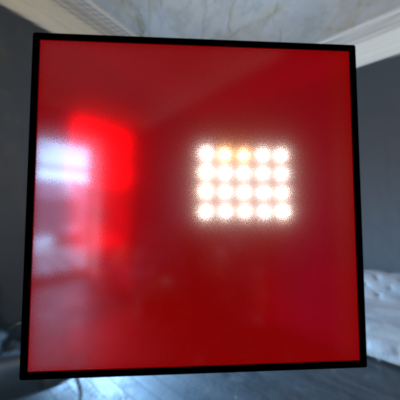} &
        \includegraphics[width=\resLen]{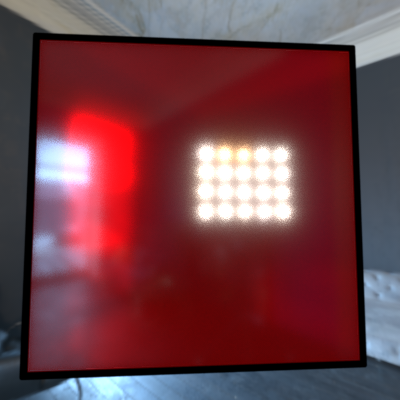} &
        \includegraphics[width=\resLen]{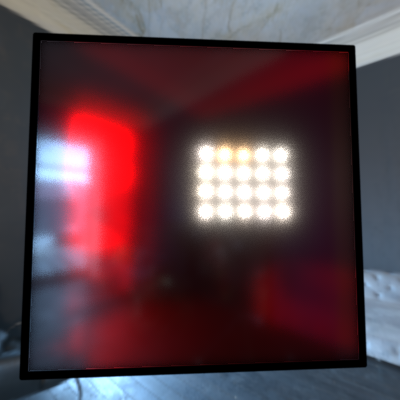} 
        \\
        & \multicolumn{6}{c}{\hspace{1.6em} (0, 0, 0) \hspace{2.2em} (0.01, 0, 0) \leftarrowfill \A \rightarrowfill (1, 0, 0) \hspace{3.4em} (1, 1, 1) \;\;\;\;\;\;}
        \\
        \raisebox{30pt}{\rotatebox[origin=c]{90}{(d) Albedo}} &
        \includegraphics[width=\resLen]{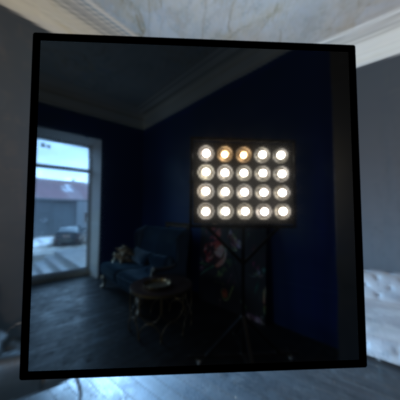} &
        \includegraphics[width=\resLen]{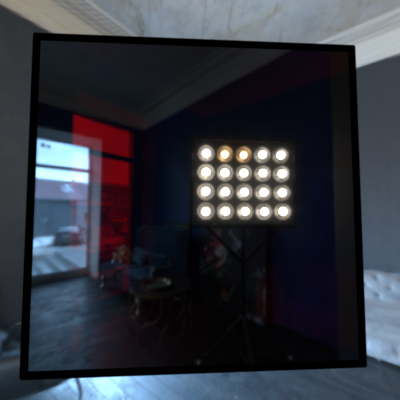} &
        \includegraphics[width=\resLen]{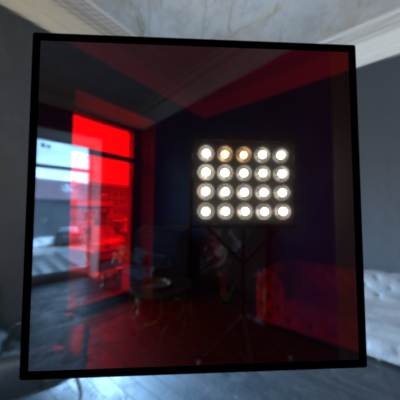} &
        \includegraphics[width=\resLen]{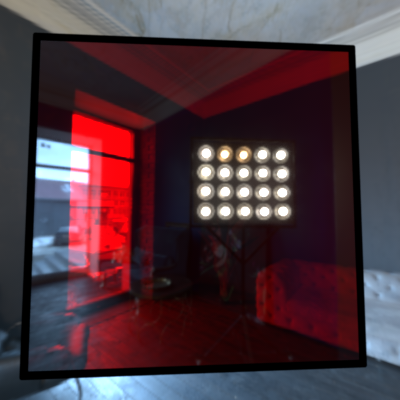} &
        \includegraphics[width=\resLen]{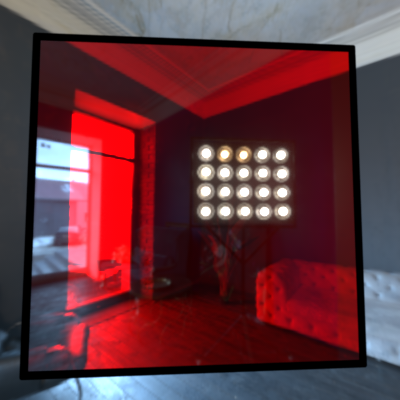} &
        \includegraphics[width=\resLen]{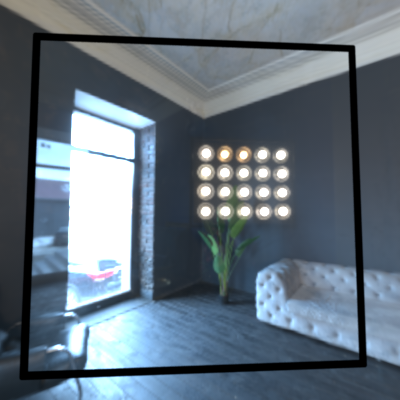} 
    \end{tabular}\vspace{-8pt}
    \caption{
        \textbf{ePBR material intrinsic evaluation}. (a) The only opaque surface in this figure, as the Metallic ($\M$) increases, the reflectance tint from light color (white) to metal color (red here); (b) Roughness ($\R$) influences how blurry both reflection and transmission appear; (c) Transparency ($\T$) controls how clear the background that can be seen; (d) Albedo ($\A$) indicates the color of the thin slab, and the background color would also be influenced.  
    }\vspace{-8pt}
    \label{fig:evaluate}
\end{figure*}

\subsection{Intrinsic evaluation}
We demonstrate how the final appearance is altered when the intrinsic properties of the material change. In \cref{fig:evaluate}, we use a flat surface illuminated by an environment map to avoid light interaction between different surfaces. With the red color of albedo, we can see the difference more clearly.

\noindent\textbf{Metallic}
With an increase in metallicity, the material transitions from non-metal to metal. The diffuse layer gradually disappears, and the highlights change from light to metal color.
Metallic changes only happen on an opaque surface. In other words, the metallic value should always be zero for a transparent or translucent surface. We can synthesize a surface that looks like a transparent metal, but to our knowledge, it is an invalid material in the real world.

\noindent\textbf{Roughness}
For a metal surface, it is more accurate to use separate roughness values and direction for the anisotropy effect. For a transparent surface such as glass, two roughness values are always used for the top and bottom surfaces \cite{guo2016rendering, guo2018position}. However, for simplicity and consistency with other materials, we use only a single roughness for all different cases. As shown in \cref{fig:evaluate}, roughness controls the blurring of both reflection and transmission. 

\noindent\textbf{Transparency}
The transparency controls how many lights can pass through the thin slab, which is the opposite of the density of the slab. As density increases (transparency decreases), more lights are scattered, making the background harder to see. The reflection on the top surface will remain unchanged.

\noindent\textbf{Albedo}
The albedo indicates nonabsorbed light. Transparency becomes ineffective with $\A=(0,0,0)$ because light cannot scatter or pass through the surface. And if $\A=(1,1,1)$, it represents pure glass with no energy loss.

\subsection{Image composition}\vspace{-5pt}
In \cref{fig:compose1}, we validate that our image composition method is more faithful to the input intrinsic channels compared to the diffusion-based method RGB$\leftrightarrow$X \cite{zeng2024rgb}. Our tested examples are from \textit{InteriorVerse} \cite{zhu2022learning} dataset, which is not contained in the training set of RGB$\leftrightarrow$X. Some of the intrinsic channels ($\N, \D, \A, \R, \M$) are directly from the ground truth. $\E$ is a Low dynamic range (LDR) image estimated using RGB$\leftrightarrow$X, $\T$ is the inverse of the ground-truth mask since we found that most of the nonmasked areas are glasses. Since $\M$ and $\T$ are not reliable in \textit{InteriorVerse}, we manually made some modifications to these channels to fit the renderings well. $\Amr$ is generated using SSRT and the out-of-space colors are set to gray. $\Abg$ is set to $1$ for the illumination behind glass.

Visually speaking, our results match well the path-traced reference regarding high-specular regions, such as floors, windows, or glass decorations on the wall.  
To avoid the color/brightness shifting caused by tone mapping, we use LPIPS \cite{zhang2018unreasonable} errors to further verify the reliability of our image composition method. 
As we can see from \cref{tab:performance}, our method performs well especially in mirror-like areas.

\renewcommand\imwidth{640}
\renewcommand\imheight{480}

\renewcommand\xone{50}
\renewcommand\yone{310}
\renewcommand\wone{150}
\renewcommand\hone{150}

\renewcommand\xtwo{260}
\renewcommand\ytwo{280}
\renewcommand\wtwo{150}
\renewcommand\htwo{150}

\renewcommand\xthr{410}
\renewcommand\ythr{280}
\renewcommand\wthr{150}
\renewcommand\hthr{150}

\renewcommand\xfou{40}
\renewcommand\yfou{140}
\renewcommand\wfou{150}
\renewcommand\hfou{150}

\renewcommand\xfiv{280}
\renewcommand\yfiv{320}
\renewcommand\wfiv{150}
\renewcommand\hfiv{150}

\renewcommand\leftone{\fpeval{\xone}}
\renewcommand\botone{\fpeval{\imheight-\yone-\hone}}
\renewcommand\rightone{\fpeval{\imwidth-\xone-\wone}}
\renewcommand\topone{\fpeval{\yone}}
\renewcommand\bxone{\fpeval{\xone/\imwidth*100}}
\renewcommand\bxxone{\fpeval{(\xone+\wone)/\imwidth*100}}
\renewcommand\byone{\fpeval{(\imheight-\yone-\hone)/\imwidth*100}}
\renewcommand\byyone{\fpeval{(\imheight-\yone)/\imwidth*100}}

\renewcommand\lefttwo{\fpeval{\xtwo}}
\renewcommand\bottwo{\fpeval{\imheight-\ytwo-\htwo}}
\renewcommand\righttwo{\fpeval{\imwidth-\xtwo-\wtwo}}
\renewcommand\toptwo{\fpeval{\ytwo}}
\renewcommand\bxtwo{\fpeval{\xtwo/\imwidth*100}}
\renewcommand\bxxtwo{\fpeval{(\xtwo+\wtwo)/\imwidth*100}}
\renewcommand\bytwo{\fpeval{(\imheight-\ytwo-\htwo)/\imwidth*100}}
\renewcommand\byytwo{\fpeval{(\imheight-\ytwo)/\imwidth*100}}

\renewcommand\leftthr{\fpeval{\xthr}}
\renewcommand\botthr{\fpeval{\imheight-\ythr-\hthr}}
\renewcommand\rightthr{\fpeval{\imwidth-\xthr-\wthr}}
\renewcommand\topthr{\fpeval{\ythr}}
\renewcommand\bxthr{\fpeval{\xthr/\imwidth*100}}
\renewcommand\bxxthr{\fpeval{(\xthr+\wthr)/\imwidth*100}}
\renewcommand\bythr{\fpeval{(\imheight-\ythr-\hthr)/\imwidth*100}}
\renewcommand\byythr{\fpeval{(\imheight-\ythr)/\imwidth*100}}

\renewcommand\leftfou{\fpeval{\xfou}}
\renewcommand\botfou{\fpeval{\imheight-\yfou-\hfou}}
\renewcommand\rightfou{\fpeval{\imwidth-\xfou-\wfou}}
\renewcommand\topfou{\fpeval{\yfou}}
\renewcommand\bxfou{\fpeval{\xfou/\imwidth*100}}
\renewcommand\bxxfou{\fpeval{(\xfou+\wfou)/\imwidth*100}}
\renewcommand\byfou{\fpeval{(\imheight-\yfou-\hfou)/\imwidth*100}}
\renewcommand\byyfou{\fpeval{(\imheight-\yfou)/\imwidth*100}}

\renewcommand\leftfiv{\fpeval{\xfiv}}
\renewcommand\botfiv{\fpeval{\imheight-\yfiv-\hfiv}}
\renewcommand\rightfiv{\fpeval{\imwidth-\xfiv-\wfiv}}
\renewcommand\topfiv{\fpeval{\yfiv}}
\renewcommand\bxfiv{\fpeval{\xfiv/\imwidth*100}}
\renewcommand\bxxfiv{\fpeval{(\xfiv+\wfiv)/\imwidth*100}}
\renewcommand\byfiv{\fpeval{(\imheight-\yfiv-\hfiv)/\imwidth*100}}
\renewcommand\byyfiv{\fpeval{(\imheight-\yfiv)/\imwidth*100}}

\begin{figure*}[t]
    \centering
    \setlength{\resLen}{0.073\linewidth}
    \setlength{\boxLen}{0.045\linewidth}
    \addtolength{\tabcolsep}{-7pt}
    \begin{tabular}{c@{\hskip 2pt}cc@{\hskip 2pt}c@{\hskip 2pt}c@{\hskip 0pt}c@{\hskip 0pt}c}
        \begin{overpic}[height=\resLen]{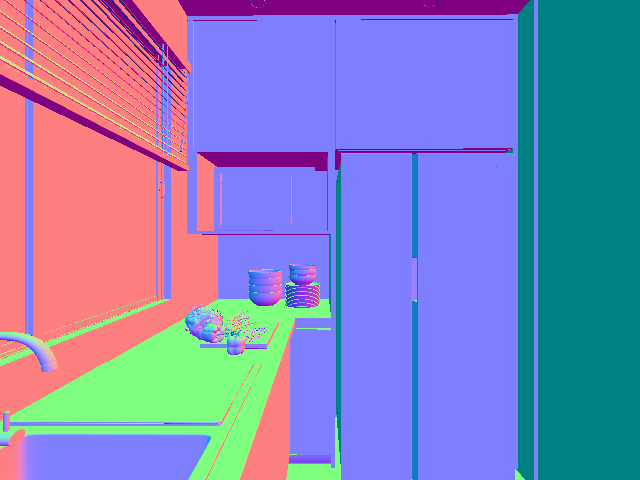} 
        \end{overpic} &
        \begin{overpic}[height=\resLen]{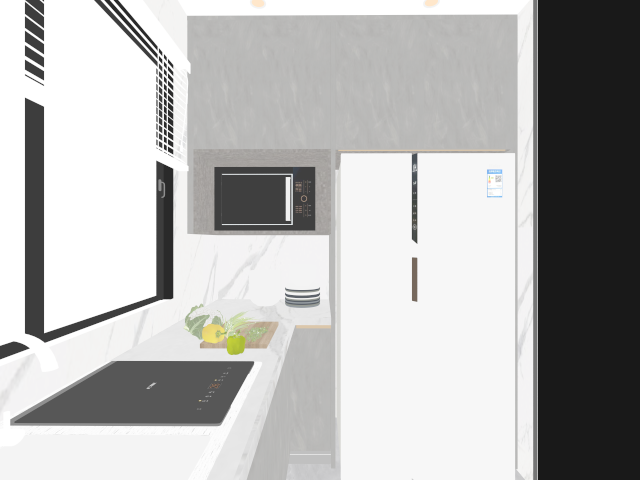} 
        \end{overpic} &
        \begin{overpic}[height=\resLen]{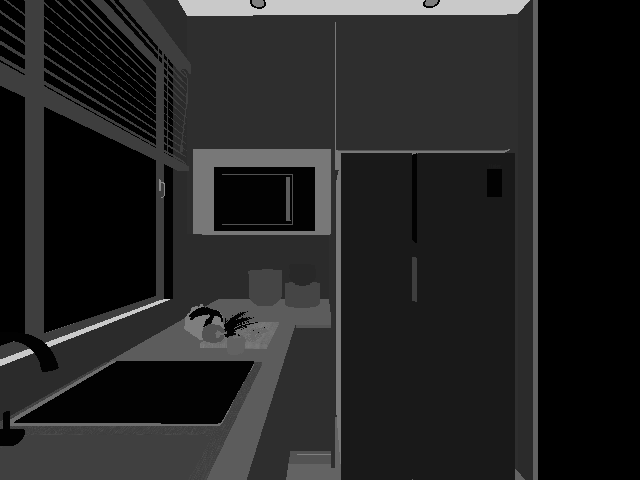} 
        \end{overpic} &
        \begin{overpic}[height=\resLen]{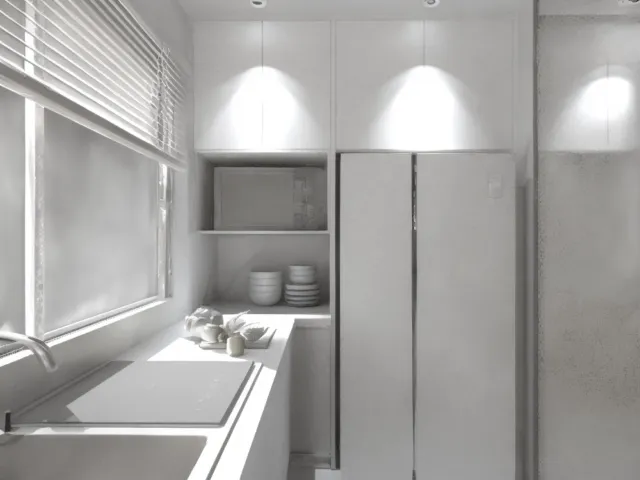} 
        \end{overpic} &
        \multirow{2}{*}[28pt]{
            \begin{overpic}[height=2\resLen]{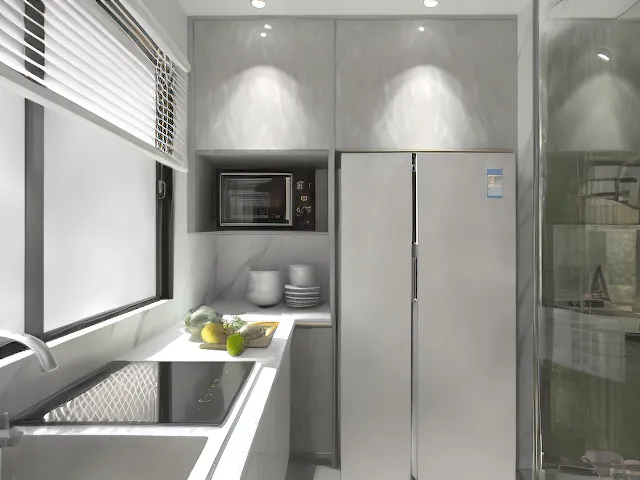} 
                \put(0,0){\color{green}\linethickness{0.5pt}
                    \polygon(\bxone,\byone)(\bxxone,\byone)(\bxxone,\byyone)(\bxone,\byyone)}
            \end{overpic}} &
        \multirow{2}{*}[28pt]{
            \begin{overpic}[height=2\resLen]{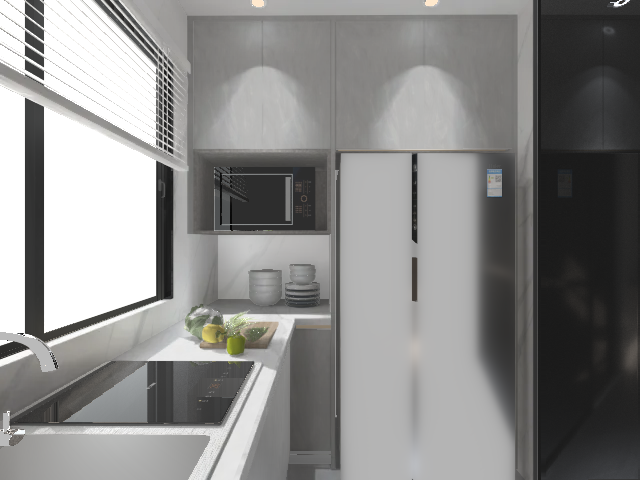} 
                \put(0,0){\color{green}\linethickness{0.5pt}
                    \polygon(\bxone,\byone)(\bxxone,\byone)(\bxxone,\byyone)(\bxone,\byyone)}
            \end{overpic}} &
        \multirow{2}{*}[28pt]{
            \begin{overpic}[height=2\resLen]{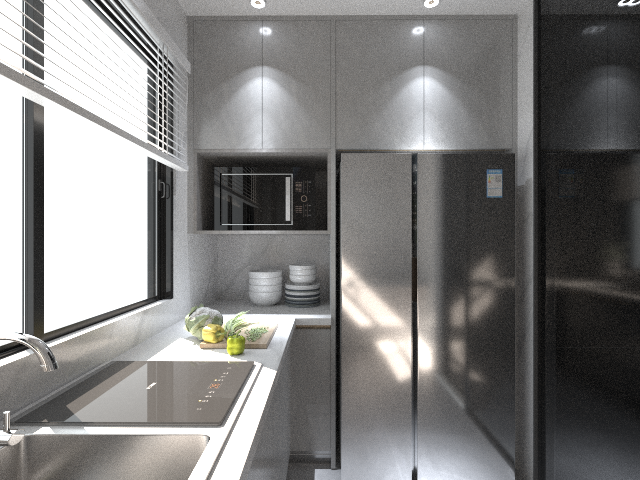} 
                \put(0,0){\color{green}\linethickness{0.5pt}
                    \polygon(\bxone,\byone)(\bxxone,\byone)(\bxxone,\byyone)(\bxone,\byyone)}
                \put(88,3){\scriptsize \colorbox{white}{\color{black} (a)}}
            \end{overpic}} \\[-4pt]
        \begin{overpic}[height=\resLen]{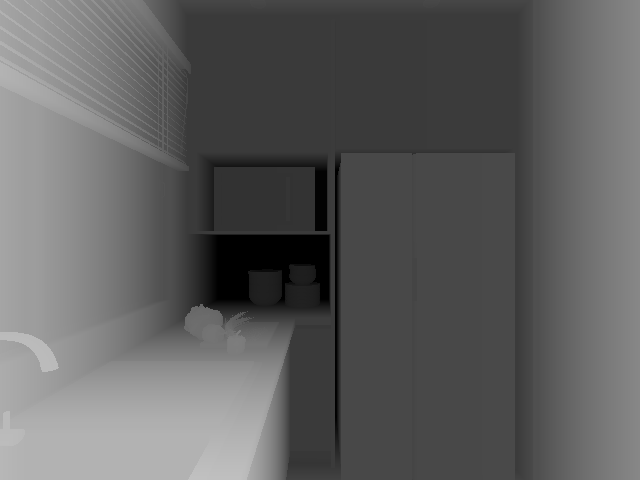} 
        \end{overpic} &
        \begin{overpic}[height=\resLen]{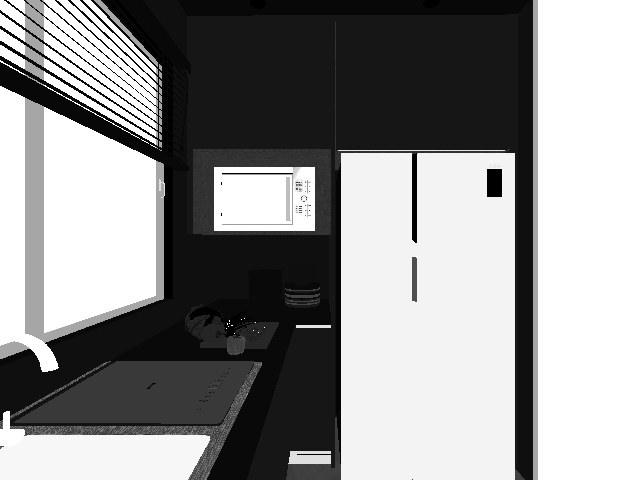} 
        \end{overpic} &
        \begin{overpic}[height=\resLen]{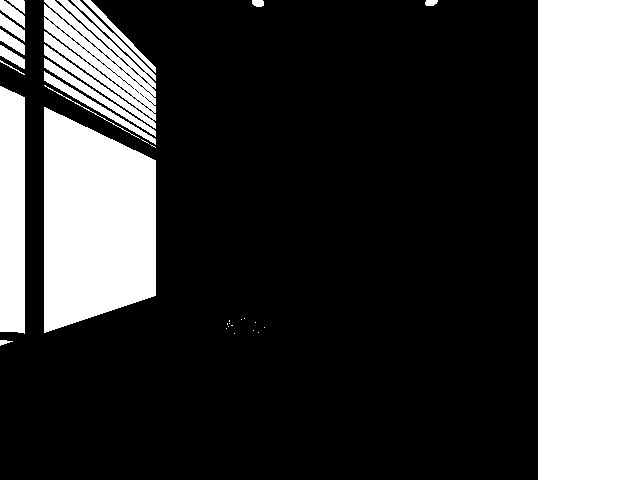} 
        \end{overpic} &
        \begin{overpic}[height=\resLen]{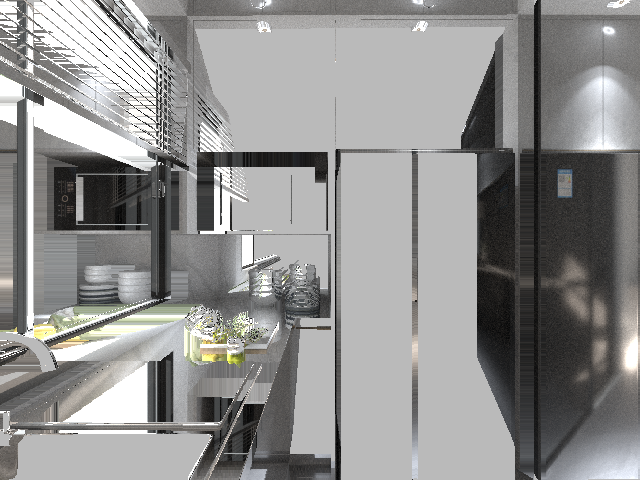} 
        \end{overpic}
        \\[-2pt]
        \begin{overpic}[height=\resLen]{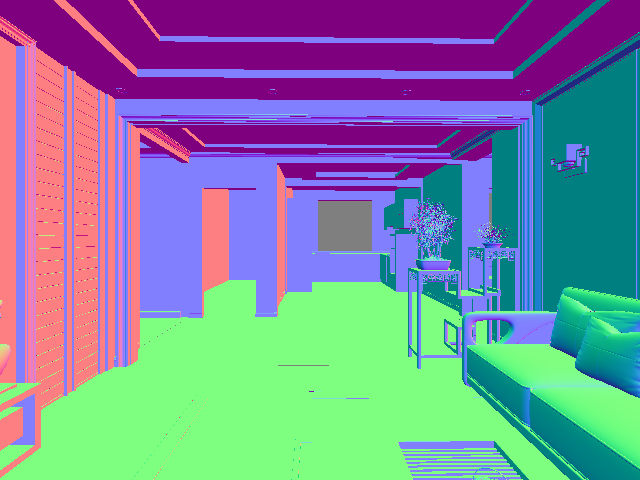} 
        \end{overpic} &
        \begin{overpic}[height=\resLen]{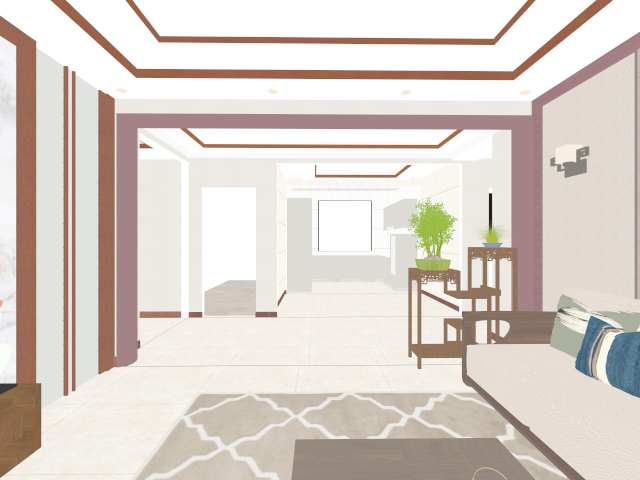} 
        \end{overpic} &
        \begin{overpic}[height=\resLen]{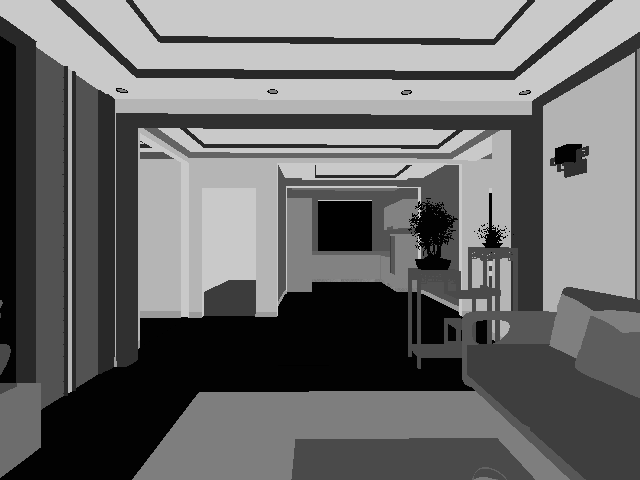} 
        \end{overpic} &
        \begin{overpic}[height=\resLen]{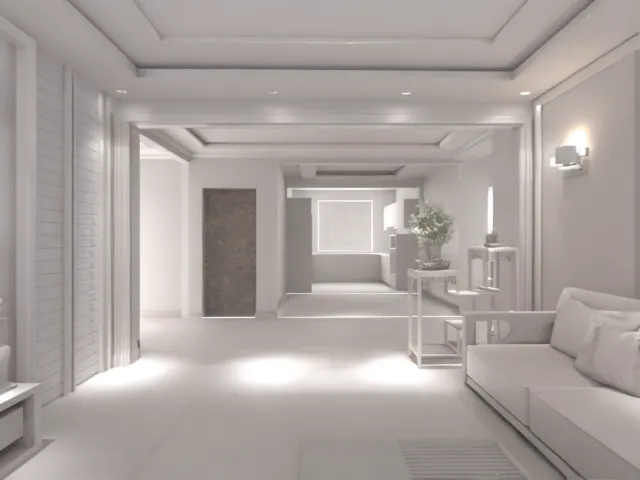} 
        \end{overpic} &
        \multirow{2}{*}[28pt]{
            \begin{overpic}[height=2\resLen]{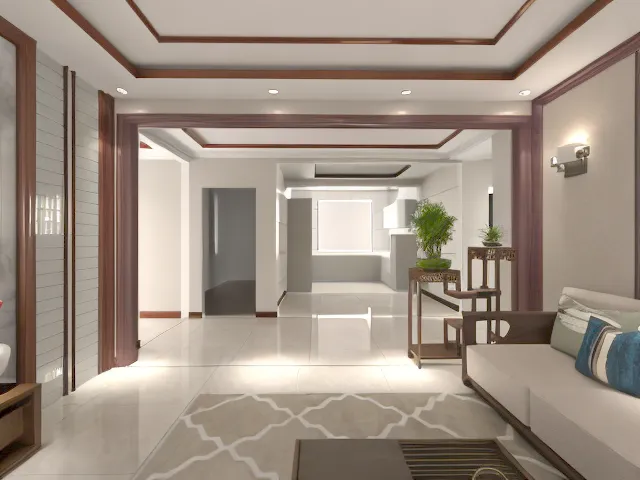} 
                \put(0,0){\color{green}\linethickness{0.5pt}
                    \polygon(\bxtwo,\bytwo)(\bxxtwo,\bytwo)(\bxxtwo,\byytwo)(\bxtwo,\byytwo)}
            \end{overpic}} &
        \multirow{2}{*}[28pt]{
            \begin{overpic}[height=2\resLen]{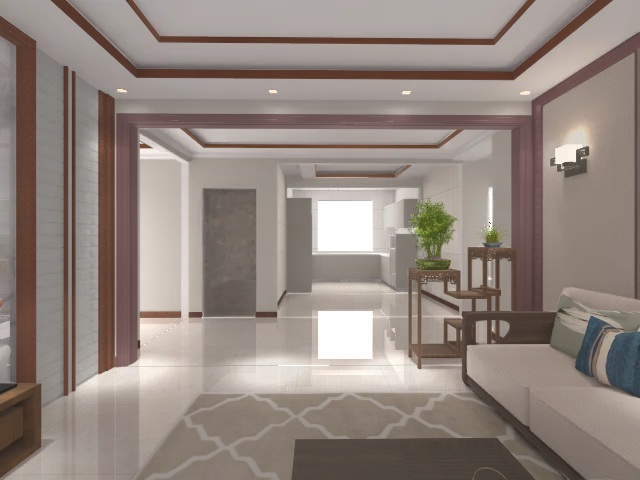} 
                \put(0,0){\color{green}\linethickness{0.5pt}
                    \polygon(\bxtwo,\bytwo)(\bxxtwo,\bytwo)(\bxxtwo,\byytwo)(\bxtwo,\byytwo)}
            \end{overpic}} &
        \multirow{2}{*}[28pt]{
            \begin{overpic}[height=2\resLen]{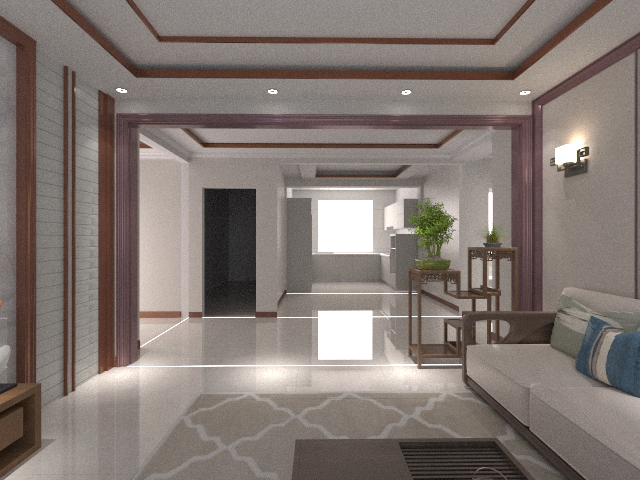} 
                \put(0,0){\color{green}\linethickness{0.5pt}
                    \polygon(\bxtwo,\bytwo)(\bxxtwo,\bytwo)(\bxxtwo,\byytwo)(\bxtwo,\byytwo)}
                \put(88,3){\scriptsize \colorbox{white}{\color{black} (b)}}
            \end{overpic}} \\[-4pt]
        \begin{overpic}[height=\resLen]{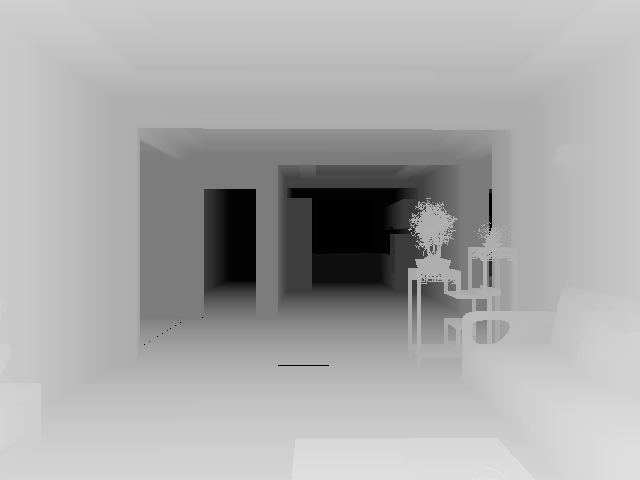} 
        \end{overpic} &
        \begin{overpic}[height=\resLen]{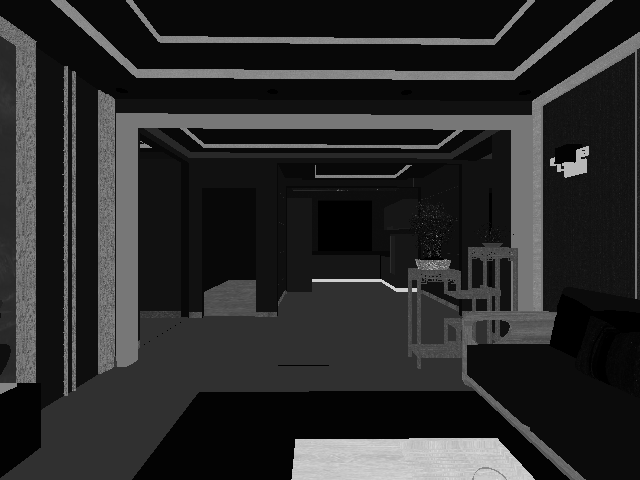} 
        \end{overpic} &
        \begin{overpic}[height=\resLen]{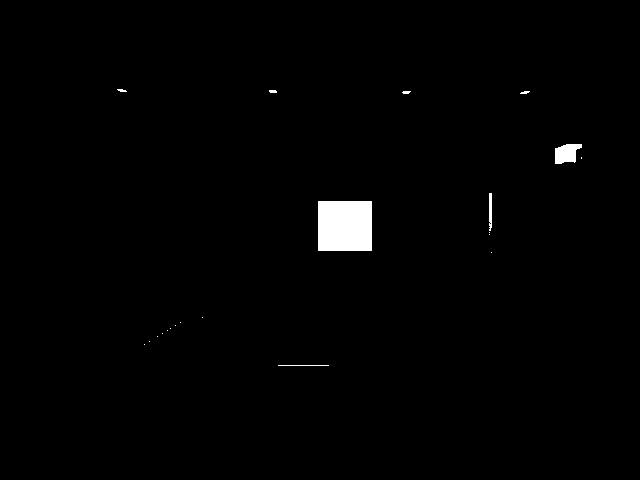} 
        \end{overpic} &
        \begin{overpic}[height=\resLen]{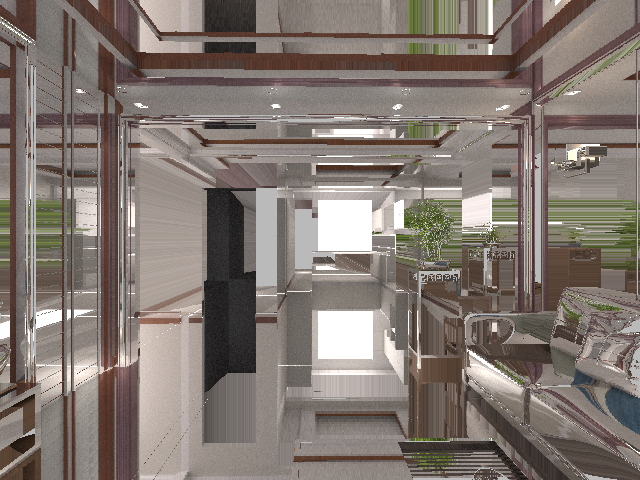} 
        \end{overpic}
        \\[-2pt]
        \begin{overpic}[height=\resLen]{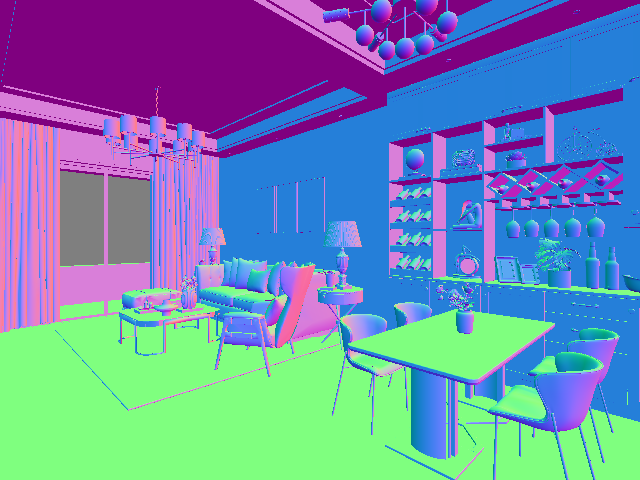} 
        \end{overpic} &
        \begin{overpic}[height=\resLen]{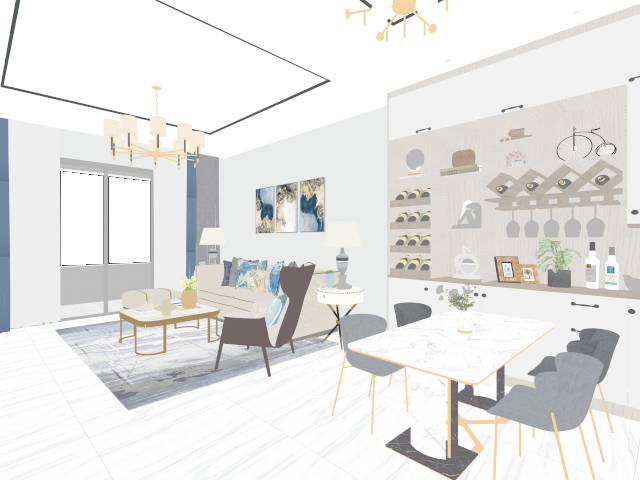} 
        \end{overpic} &
        \begin{overpic}[height=\resLen]{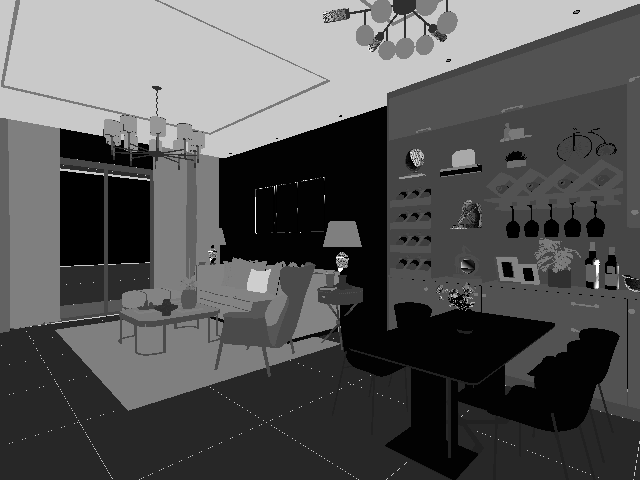} 
        \end{overpic} &
        \begin{overpic}[height=\resLen]{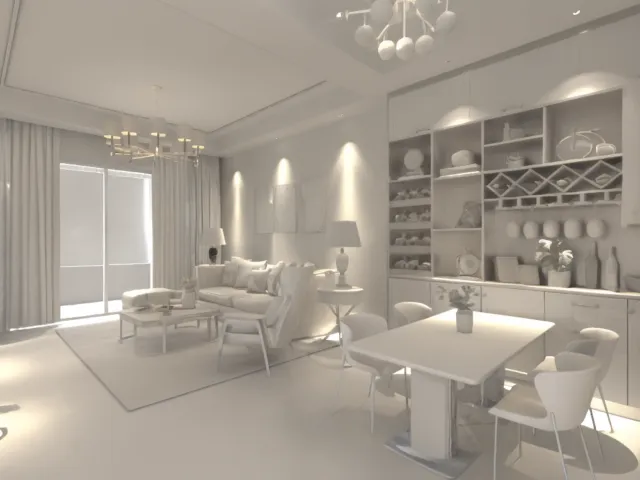} 
        \end{overpic} &
        \multirow{2}{*}[28pt]{
            \begin{overpic}[height=2\resLen]{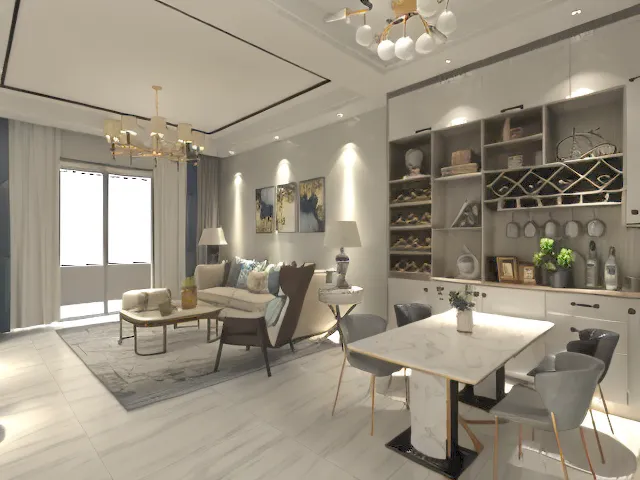} 
                \put(0,0){\color{green}\linethickness{0.5pt}
                    \polygon(\bxthr,\bythr)(\bxxthr,\bythr)(\bxxthr,\byythr)(\bxthr,\byythr)}
            \end{overpic}} &
        \multirow{2}{*}[28pt]{
            \begin{overpic}[height=2\resLen]{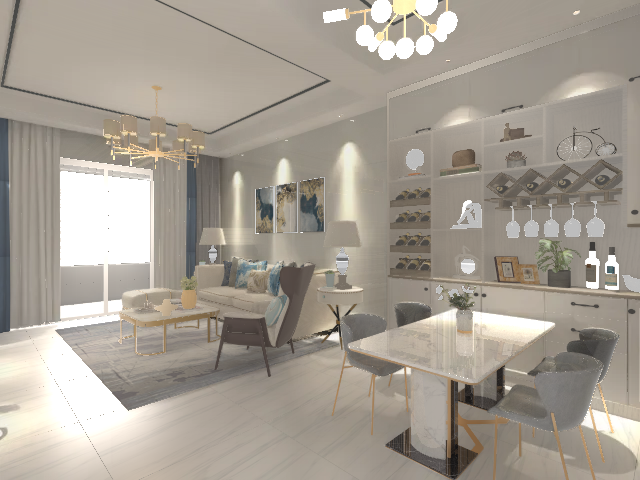} 
                \put(0,0){\color{green}\linethickness{0.5pt}
                    \polygon(\bxthr,\bythr)(\bxxthr,\bythr)(\bxxthr,\byythr)(\bxthr,\byythr)}
            \end{overpic}} &
        \multirow{2}{*}[28pt]{
            \begin{overpic}[height=2\resLen]{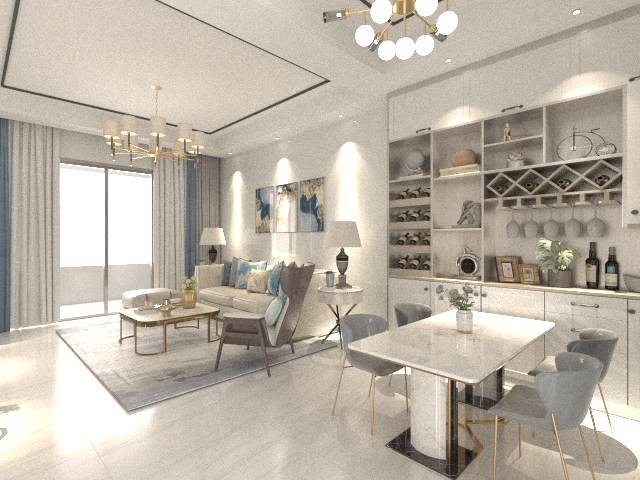} 
                \put(0,0){\color{green}\linethickness{0.5pt}
                    \polygon(\bxthr,\bythr)(\bxxthr,\bythr)(\bxxthr,\byythr)(\bxthr,\byythr)}
                \put(88,3){\scriptsize \colorbox{white}{\color{black} (c)}}
            \end{overpic}} \\[-4pt]
        \begin{overpic}[height=\resLen]{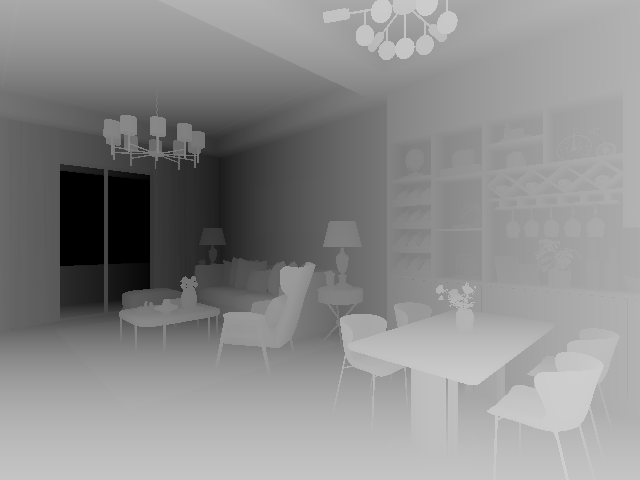} 
        \end{overpic} &
        \begin{overpic}[height=\resLen]{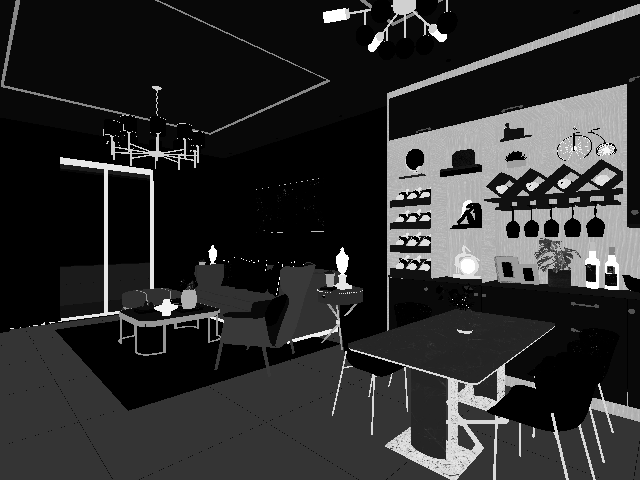} 
        \end{overpic} &
        \begin{overpic}[height=\resLen]{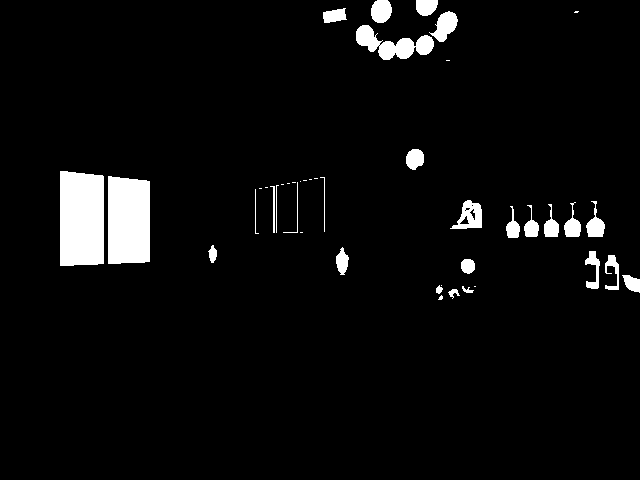} 
        \end{overpic} &
        \begin{overpic}[height=\resLen]{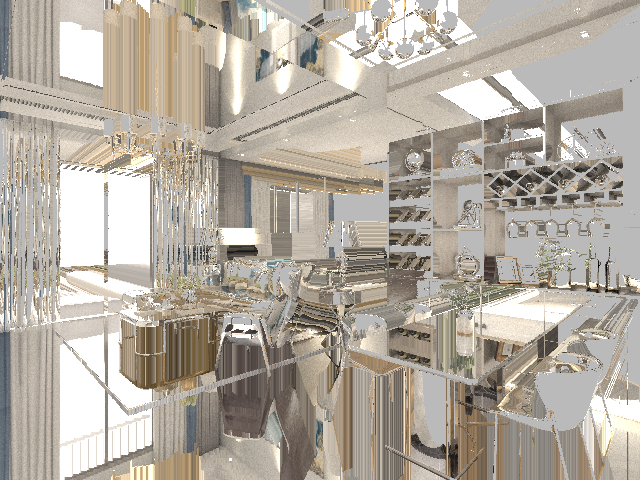} 
        \end{overpic}
        \\[-2pt]
        \begin{overpic}[height=\resLen]{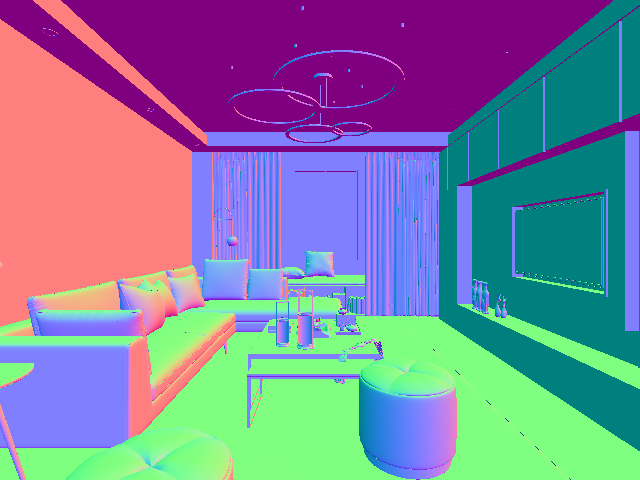} 
        \end{overpic} &
        \begin{overpic}[height=\resLen]{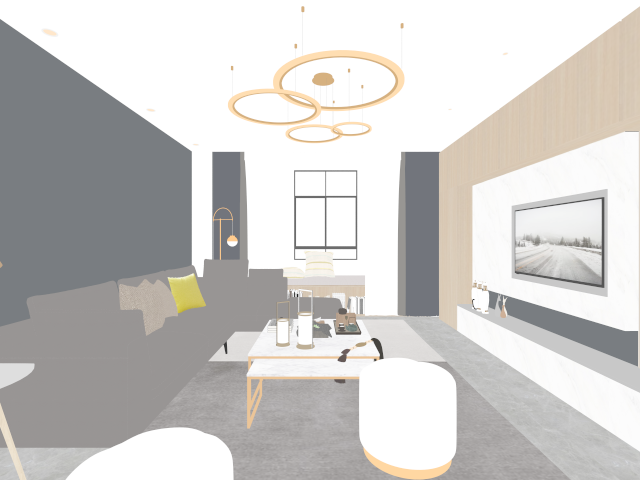} 
        \end{overpic} &
        \begin{overpic}[height=\resLen]{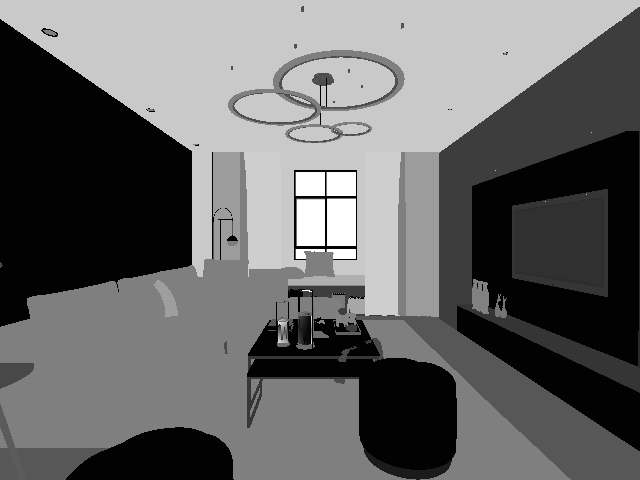} 
        \end{overpic} &
        \begin{overpic}[height=\resLen]{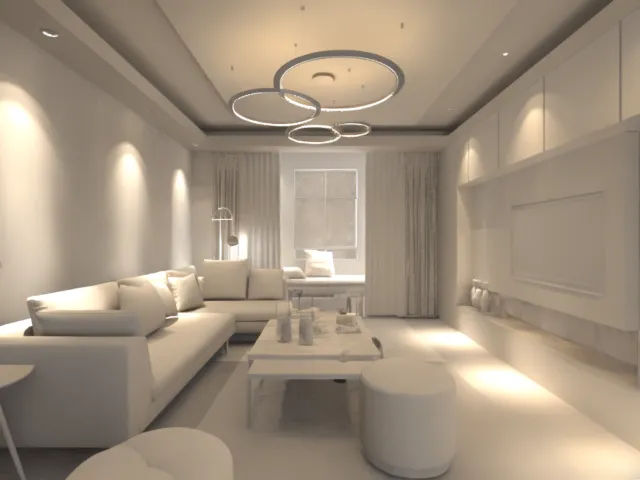} 
        \end{overpic} &
        \multirow{2}{*}[28pt]{
            \begin{overpic}[height=2\resLen]{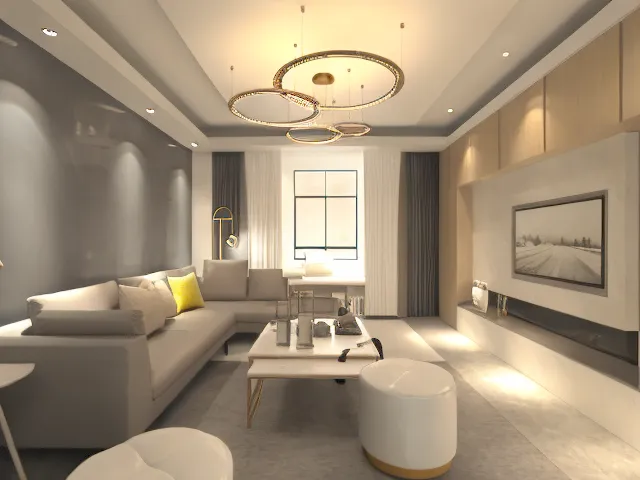} 
                \put(0,0){\color{green}\linethickness{0.5pt}
                    \polygon(\bxfou,\byfou)(\bxxfou,\byfou)(\bxxfou,\byyfou)(\bxfou,\byyfou)}
            \end{overpic}} &
        \multirow{2}{*}[28pt]{
            \begin{overpic}[height=2\resLen]{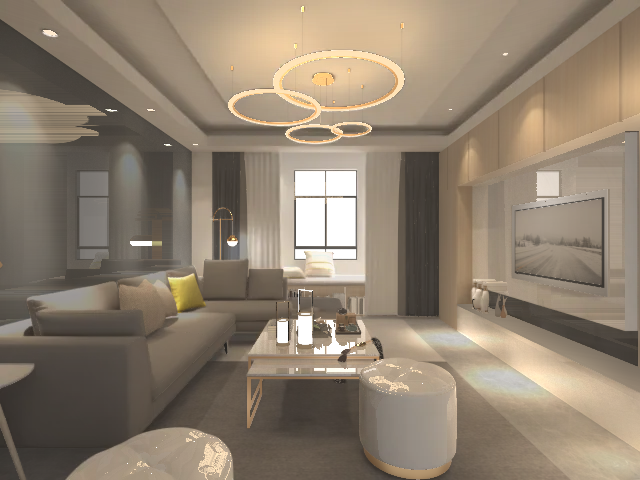} 
                \put(0,0){\color{green}\linethickness{0.5pt}
                    \polygon(\bxfou,\byfou)(\bxxfou,\byfou)(\bxxfou,\byyfou)(\bxfou,\byyfou)}
            \end{overpic}} &
        \multirow{2}{*}[28pt]{
            \begin{overpic}[height=2\resLen]{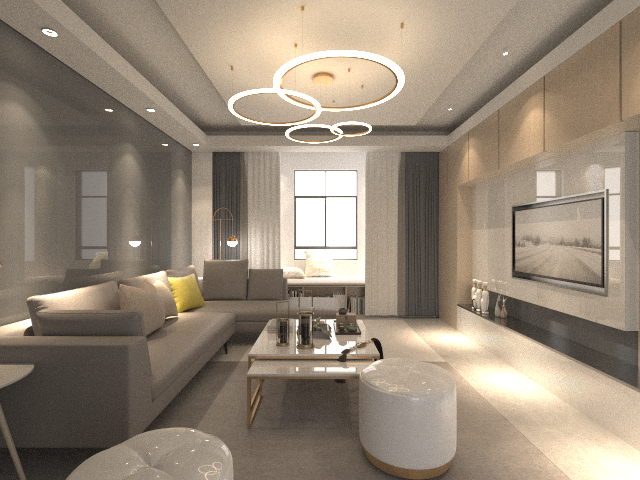} 
                \put(0,0){\color{green}\linethickness{0.5pt}
                    \polygon(\bxfou,\byfou)(\bxxfou,\byfou)(\bxxfou,\byyfou)(\bxfou,\byyfou)}
                \put(88,3){\scriptsize \colorbox{white}{\color{black} (d)}}
            \end{overpic}} \\[-4pt]
        \begin{overpic}[height=\resLen]{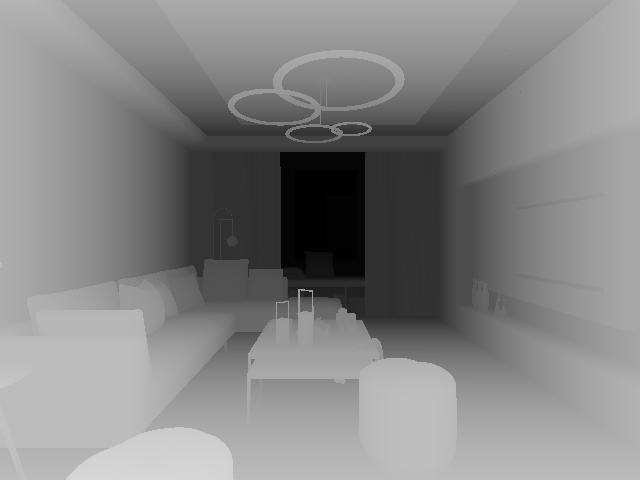} 
        \end{overpic} &
        \begin{overpic}[height=\resLen]{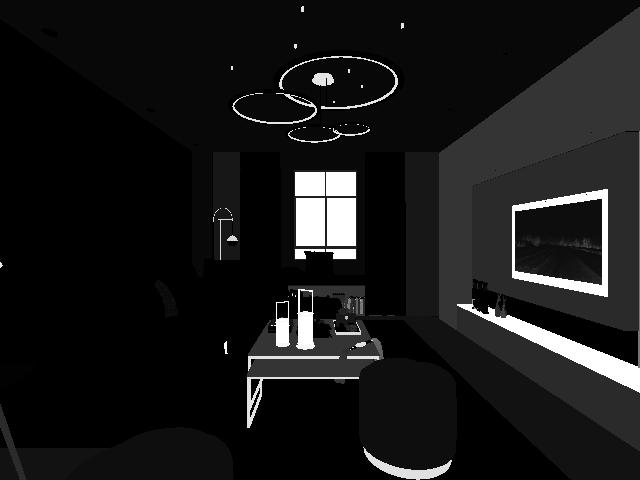} 
        \end{overpic} &
        \begin{overpic}[height=\resLen]{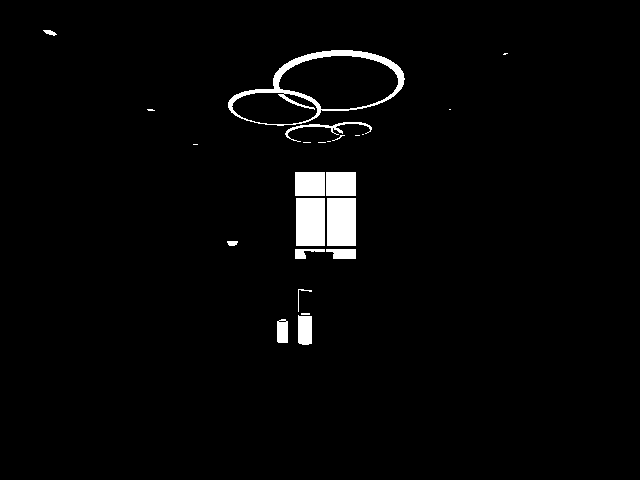} 
        \end{overpic} &
        \begin{overpic}[height=\resLen]{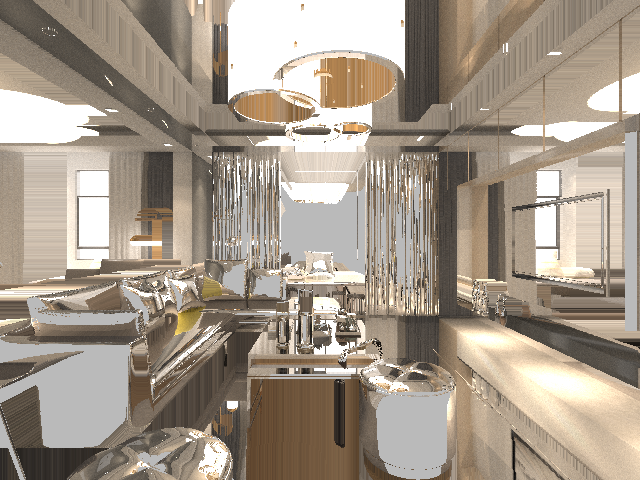} 
        \end{overpic}
        \\[-2pt]
        \begin{overpic}[height=\resLen]{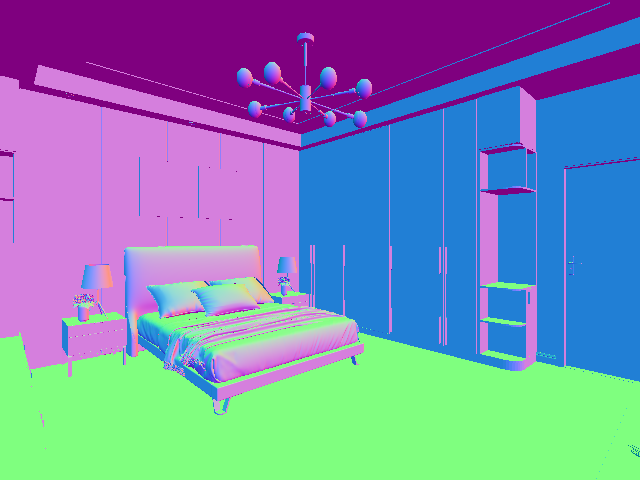} 
        \end{overpic} &
        \begin{overpic}[height=\resLen]{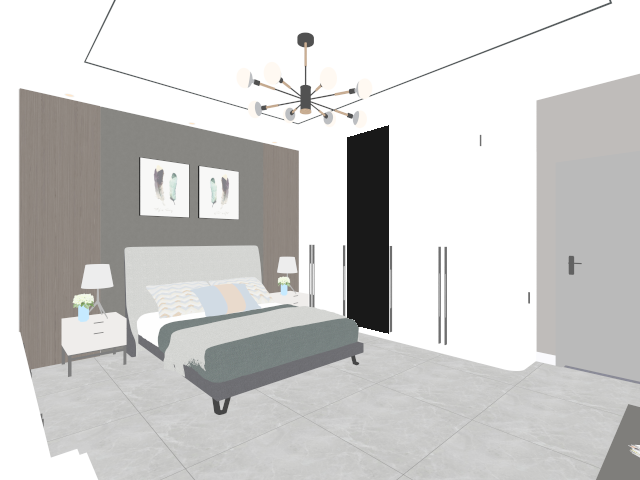} 
        \end{overpic} &
        \begin{overpic}[height=\resLen]{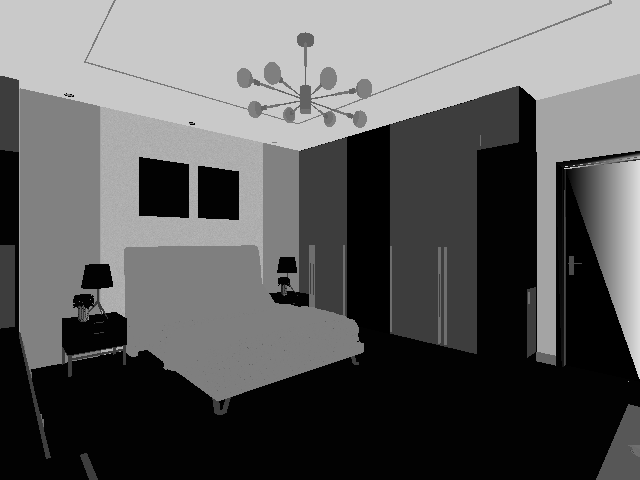} 
        \end{overpic} &
        \begin{overpic}[height=\resLen]{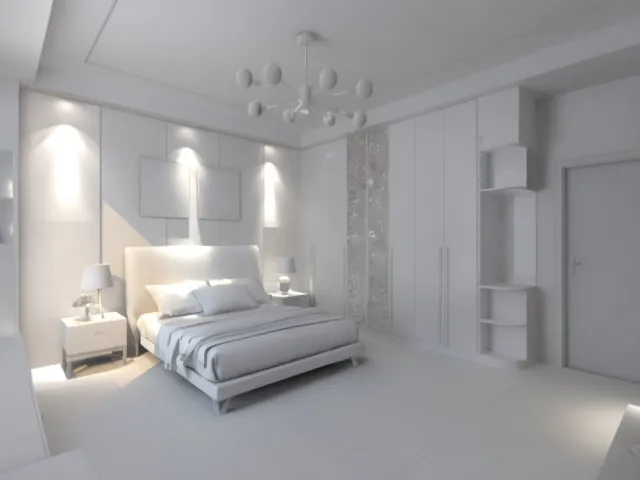} 
        \end{overpic} &
        \multirow{2}{*}[28pt]{
            \begin{overpic}[height=2\resLen]{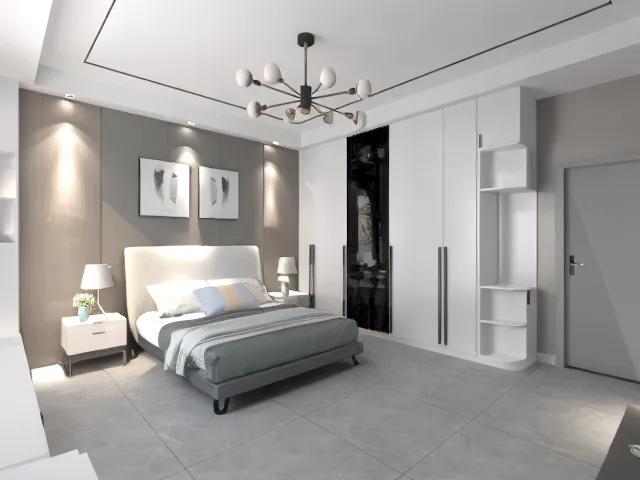} 
                \put(0,0){\color{green}\linethickness{0.5pt}
                    \polygon(\bxfiv,\byfiv)(\bxxfiv,\byfiv)(\bxxfiv,\byyfiv)(\bxfiv,\byyfiv)}
            \end{overpic}} &
        \multirow{2}{*}[28pt]{
            \begin{overpic}[height=2\resLen]{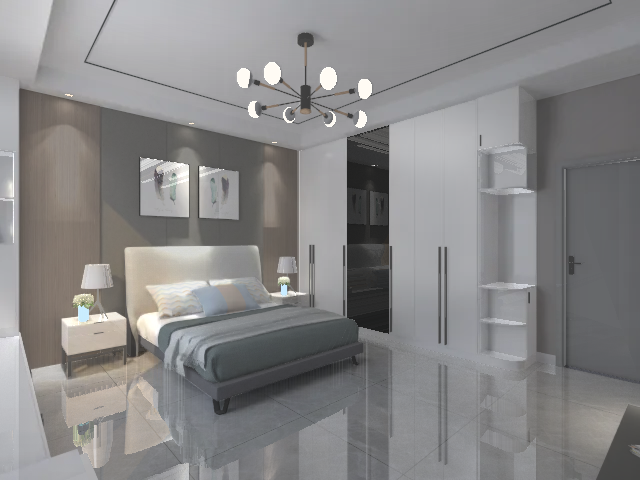} 
                \put(0,0){\color{green}\linethickness{0.5pt}
                    \polygon(\bxfiv,\byfiv)(\bxxfiv,\byfiv)(\bxxfiv,\byyfiv)(\bxfiv,\byyfiv)}
            \end{overpic}} &
        \multirow{2}{*}[28pt]{
            \begin{overpic}[height=2\resLen]{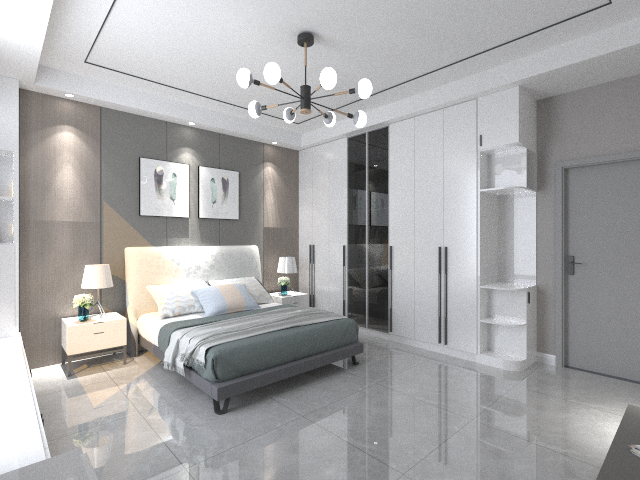} 
                \put(0,0){\color{green}\linethickness{0.5pt}
                    \polygon(\bxfiv,\byfiv)(\bxxfiv,\byfiv)(\bxxfiv,\byyfiv)(\bxfiv,\byyfiv)}
                \put(88,3){\scriptsize \colorbox{white}{\color{black} (e)}}
            \end{overpic}} \\[-4pt]
        \begin{overpic}[height=\resLen]{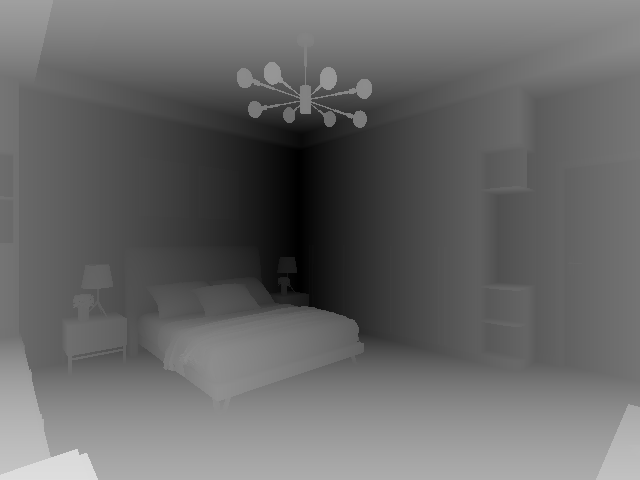} 
        \end{overpic} &
        \begin{overpic}[height=\resLen]{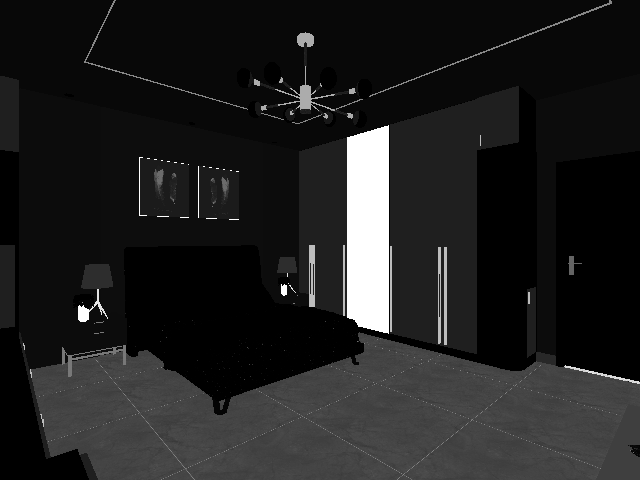} 
        \end{overpic} &
        \begin{overpic}[height=\resLen]{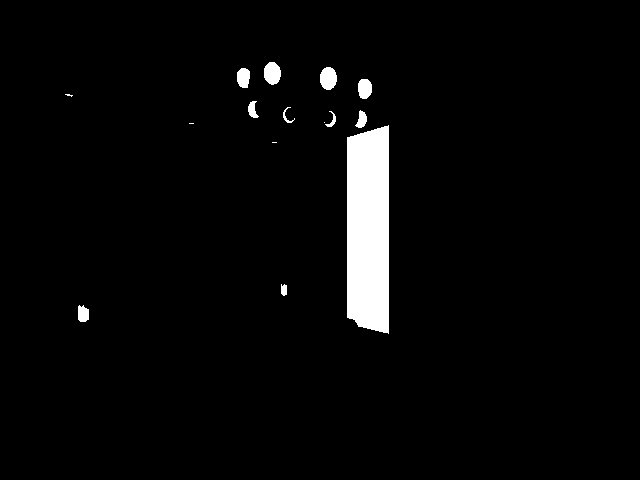} 
        \end{overpic} &
        \begin{overpic}[height=\resLen]{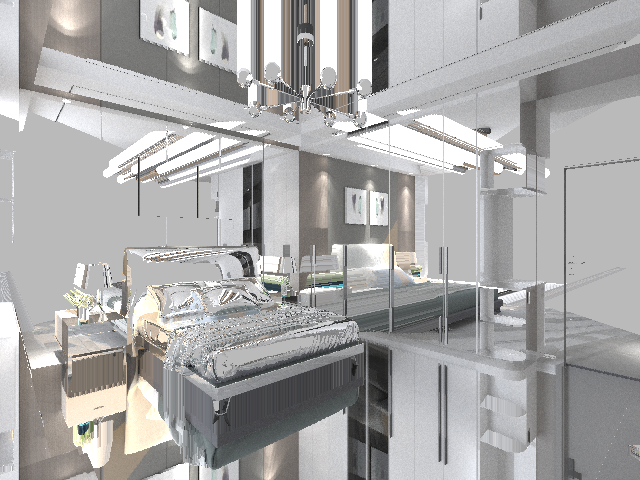} 
        \end{overpic}
        \\[-4pt] 
        \footnotesize{$\N$ and $\D$} & \multicolumn{2}{c}{\footnotesize{$\A$,$\R$,$\M$ and $\T$}} & \footnotesize{$\E$ and $\Amr$} & \multirow{2}{*}{RGB$\leftrightarrow$X \cite{zeng2024rgb}} & \multirow{2}{*}{Ours} & \multirow{2}{*}{Reference} \\
        \multicolumn{4}{c}{Intrinsic channels}
    \end{tabular}
    \caption{
        \textbf{Image composition.} We decompose reference images into intrinsic channels and then recompose them back. Compared to RGB$\leftrightarrow$X \cite{zeng2024rgb}, our results generate more accurate reflectance in the high specular regions (marked with green box).
        In some non-specular regions, our light diffusion is less bright compared to reference, which is mainly because of the inaccurate LDR irradiance map $\E$. 
    }
    \label{fig:compose1}
\end{figure*}



\section{Discussion}

\textit{PBR materials} is a conventionally accepted concept which is not equal to \textit{PBR}. The latter includes everything in rendering, materials, light transport, etc. \textit{PBR materials} is not actually physically based, and are usually referred to the reflection model in real-time rendering. Our \textit{ePBR materials} is still a simplified model without considering anisotropic and subsurface scattering. Even for a transparent object, only \textit{thin surface} is supported in our model. As shown in \cref{fig:gallery} (bottom right), to simulate accurate refraction, we need the exact geometry of the backside of the object, which is hard to estimate from knowledge of the screen space.  

If there is a dataset generated based on our \textit{ePBR materials}, all the learning-based methods that used \textit{PBR materials}, e.g., \textit{diffusion}-based image decomposition, image synthesis or texture generation, could be easily extended and improved.  

Our image composition method is relatively straightforward and can not handle multireflection or color bleeding. The composed image is entirely dependent on the accuracy of the intrinsic channels. 

\begin{table}[b]
    \centering
    \addtolength{\tabcolsep}{-2pt}
    \begin{tabular}{cccccc}
        \toprule
         & \multicolumn{5}{c}{LPIPS$\downarrow$} \\[1pt] 
         \cline{2-6}  
         & \cref{fig:compose1}(a) & (b) & (c) & (d) & (e) \\
        \midrule
        RGB$\leftrightarrow$X & $0.493$ & $0.360$ & $0.372$ & $0.387$ & $0.327$ \\
        Ours & $\bm{0.414}$ & $\bm{0.271}$ & $\bm{0.360}$ & $\bm{0.336}$ & $\bm{0.304}$ \\
        \bottomrule
    \end{tabular}
    \caption{
        \textbf{Overall image composition performance in LPIPS for \cref{fig:compose1}}.
        It shows that our composition method outperforms \textit{diffusion}-based method for the entire image.}
    \label{tab:performance}
\end{table}

\section{Conclusion}

In this work, we extended intrinsic representations to incorporate both reflection and transmission, enabling the synthesis of transparent materials such as glass and windows. By introducing an explicit intrinsic compositing framework, we achieved deterministic and interpretable image synthesis that offers precise control over material properties. Compared to \textit{diffusion}-based rendering methods, our approach provides an efficient and memory-friendly solution, making it well suited for real-time applications and high-resolution image generation. Furthermore, the image composition using our \textit{extended PBR materials} allows flexible material editing while maintaining physical plausibility. Future work will focus on exploring the integration of \textit{ePBR materials} into more 3D vision applications, bridging the gap between physically based and AI-driven image synthesis. We believe that our work can further impact downstream applications and allow for improved control over low-level properties of objects.

{
    \small
    \bibliographystyle{ieeenat_fullname}
    \bibliography{main}

\begin{thebibliography}{57}
\providecommand{\natexlab}[1]{#1}
\providecommand{\url}[1]{\texttt{#1}}
\expandafter\ifx\csname urlstyle\endcsname\relax
  \providecommand{\doi}[1]{doi: #1}\else
  \providecommand{\doi}{doi: \begingroup \urlstyle{rm}\Url}\fi

\bibitem[Belcour(2018)]{belcour2018efficient}
Laurent Belcour.
\newblock Efficient rendering of layered materials using an atomic decomposition with statistical operators.
\newblock \emph{ACM TOG}, 37\penalty0 (4):\penalty0 1, 2018.

\bibitem[Belcour and Barla(2017)]{belcour2017practical}
Laurent Belcour and Pascal Barla.
\newblock A practical extension to microfacet theory for the modeling of varying iridescence.
\newblock \emph{ACM TOG}, 36\penalty0 (4):\penalty0 1--14, 2017.

\bibitem[Blender()]{blender}
Blender.
\newblock Blender 4.3 manual: Principled {BSDF}.

\bibitem[Burley(2015)]{burley2015extending}
Brent Burley.
\newblock Extending the {D}isney {BRDF} to a {BSDF} with integrated subsurface scattering.
\newblock In \emph{ACM SIGGRAPH Course: Physically Based Shading in Theory and Practice}, page~9, 2015.

\bibitem[Burley and Studios(2012)]{burley2012physically}
Brent Burley and Walt Disney~Animation Studios.
\newblock Physically-based shading at {D}isney.
\newblock In \emph{ACM SIGGRAPH Course: Practical Physically Based Shading in Film and Game Production}, pages 1--7. vol. 2012, 2012.

\bibitem[Careaga and Aksoy(2023)]{careaga2023intrinsic}
Chris Careaga and Ya{\u{g}}{\i}z Aksoy.
\newblock Intrinsic image decomposition via ordinal shading.
\newblock \emph{ACM TOG}, 43\penalty0 (1):\penalty0 1--24, 2023.

\bibitem[Careaga and Aksoy(2024)]{careaga2024colorful}
Chris Careaga and Ya{\u{g}}{\i}z Aksoy.
\newblock Colorful diffuse intrinsic image decomposition in the wild.
\newblock \emph{ACM TOG}, 43\penalty0 (6):\penalty0 1--12, 2024.

\bibitem[Chiang et~al.(2015)Chiang, Bitterli, Tappan, and Burley]{chiang2015practical}
Matt Jen-Yuan Chiang, Benedikt Bitterli, Chuck Tappan, and Brent Burley.
\newblock A practical and controllable hair and fur model for production path tracing.
\newblock In \emph{ACM SIGGRAPH Talks}, pages 1--1. 2015.

\bibitem[Cook and Torrance(1982)]{cook1982reflectance}
Robert~L Cook and Kenneth~E. Torrance.
\newblock A reflectance model for computer graphics.
\newblock \emph{ACM TOG}, 1\penalty0 (1):\penalty0 7--24, 1982.

\bibitem[Dai et~al.(2009)Dai, Wang, Liu, Snyder, Wu, and Guo]{dai2009dual}
Qiang Dai, Jiaping Wang, Yiming Liu, John Snyder, Enhua Wu, and Baining Guo.
\newblock The dual-microfacet model for capturing thin transparent slabs.
\newblock \emph{Computer Graphics Forum}, 28\penalty0 (7):\penalty0 1917--1925, 2009.

\bibitem[Deschaintre et~al.(2018)Deschaintre, Aittala, Durand, Drettakis, and Bousseau]{deschaintre2018single}
Valentin Deschaintre, Miika Aittala, Fredo Durand, George Drettakis, and Adrien Bousseau.
\newblock Single-image {SVBRDF} capture with a rendering-aware deep network.
\newblock \emph{ACM TOG}, 37\penalty0 (4):\penalty0 1--15, 2018.

\bibitem[Guill{\'e}n et~al.(2020)Guill{\'e}n, Marco, Gutierrez, Jakob, and Jarabo]{guillen2020general}
Ib{\'o}n Guill{\'e}n, Julio Marco, Diego Gutierrez, Wenzel Jakob, and Adrian Jarabo.
\newblock A general framework for pearlescent materials.
\newblock \emph{ACM TOG}, 39\penalty0 (6):\penalty0 1--15, 2020.

\bibitem[Guo et~al.(2016)Guo, Qian, Guo, and Pan]{guo2016rendering}
Jie Guo, Jinghui Qian, Yanwen Guo, and Jingui Pan.
\newblock Rendering thin transparent layers with extended normal distribution functions.
\newblock \emph{IEEE TVCG}, 23\penalty0 (9):\penalty0 2108--2119, 2016.

\bibitem[Guo et~al.(2018)Guo, Ha{\v{s}}an, and Zhao]{guo2018position}
Yu Guo, Milo{\v{s}} Ha{\v{s}}an, and Shuang Zhao.
\newblock Position-free {M}onte {C}arlo simulation for arbitrary layered {BSDF}s.
\newblock \emph{ACM TOG}, 37\penalty0 (6):\penalty0 1--14, 2018.

\bibitem[Guo et~al.(2020)Guo, Smith, Ha{\v{s}}an, Sunkavalli, and Zhao]{guo2020materialgan}
Yu Guo, Cameron Smith, Milo{\v{s}} Ha{\v{s}}an, Kalyan Sunkavalli, and Shuang Zhao.
\newblock Material{GAN}: Reflectance capture using a generative {SVBRDF} model.
\newblock \emph{ACM TOG}, 39\penalty0 (6):\penalty0 1--13, 2020.

\bibitem[He et~al.(2025)He, Li, Yin, Liang, Li, Zhou, Zhang, Liu, and Chen]{he2024lotus}
Jing He, Haodong Li, Wei Yin, Yixun Liang, Leheng Li, Kaiqiang Zhou, Hongbo Zhang, Bingbing Liu, and Ying-Cong Chen.
\newblock Lotus: Diffusion-based visual foundation model for high-quality dense prediction.
\newblock In \emph{ICLR}, 2025.

\bibitem[Huang et~al.(2025{\natexlab{a}})Huang, Wang, Liu, and Wang]{huang2024material}
Xin Huang, Tengfei Wang, Ziwei Liu, and Qing Wang.
\newblock Material anything: Generating materials for any 3d object via diffusion.
\newblock In \emph{CVPR}, 2025{\natexlab{a}}.

\bibitem[Huang et~al.(2025{\natexlab{b}})Huang, Huang, Liu, Yan, Lv, Liu, Xiong, Zhang, Cao, and Chen]{huang2025diffusion}
Yi Huang, Jiancheng Huang, Yifan Liu, Mingfu Yan, Jiaxi Lv, Jianzhuang Liu, Wei Xiong, He Zhang, Liangliang Cao, and Shifeng Chen.
\newblock Diffusion model-based image editing: A survey.
\newblock \emph{IEEE TPAMI}, \penalty0 (01):\penalty0 1--27, 2025{\natexlab{b}}.

\bibitem[Jakob et~al.(2014)Jakob, d'Eon, Jakob, and Marschner]{jakob2014comprehensive}
Wenzel Jakob, Eugene d'Eon, Otto Jakob, and Steve Marschner.
\newblock A comprehensive framework for rendering layered materials.
\newblock \emph{ACM TOG}, 33\penalty0 (4):\penalty0 1--14, 2014.

\bibitem[Jakob et~al.(2022)Jakob, Speierer, Roussel, Nimier-David, Vicini, Zeltner, Nicolet, Crespo, Leroy, and Zhang]{jakob2022mitsuba3}
Wenzel Jakob, Sébastien Speierer, Nicolas Roussel, Merlin Nimier-David, Delio Vicini, Tizian Zeltner, Baptiste Nicolet, Miguel Crespo, Vincent Leroy, and Ziyi Zhang.
\newblock Mitsuba 3 renderer, 2022.

\bibitem[Karis and Games(2013)]{karis2013real}
Brian Karis and Epic Games.
\newblock Real shading in {U}nreal {E}ngine 4.
\newblock \emph{Proc. Physically Based Shading Theory Practice}, 4\penalty0 (3):\penalty0 1, 2013.

\bibitem[Kocsis et~al.(2024)Kocsis, Sitzmann, and Nie{\ss}ner]{kocsis2024intrinsic}
Peter Kocsis, Vincent Sitzmann, and Matthias Nie{\ss}ner.
\newblock Intrinsic image diffusion for indoor single-view material estimation.
\newblock In \emph{CVPR}, pages 5198--5208, 2024.

\bibitem[Li et~al.(2020)Li, Shafiei, Ramamoorthi, Sunkavalli, and Chandraker]{li2020inverse}
Zhengqin Li, Mohammad Shafiei, Ravi Ramamoorthi, Kalyan Sunkavalli, and Manmohan Chandraker.
\newblock Inverse rendering for complex indoor scenes: Shape, spatially-varying lighting and {SVBRDF} from a single image.
\newblock In \emph{CVPR}, pages 2475--2484, 2020.

\bibitem[Liang et~al.(2025)Liang, Gojcic, Ling, Munkberg, Hasselgren, Lin, Gao, Keller, Vijaykumar, Fidler, et~al.]{liang2025diffusionrenderer}
Ruofan Liang, Zan Gojcic, Huan Ling, Jacob Munkberg, Jon Hasselgren, Zhi-Hao Lin, Jun Gao, Alexander Keller, Nandita Vijaykumar, Sanja Fidler, et~al.
\newblock Diffusion{R}enderer: Neural inverse and forward rendering with video diffusion models.
\newblock In \emph{CVPR}, 2025.

\bibitem[Luo et~al.(2024)Luo, Ceylan, Yoon, Zhao, Philip, Fr{\"u}hst{\"u}ck, Li, Richardt, and Wang]{luo2024intrinsicdiffusion}
Jundan Luo, Duygu Ceylan, Jae~Shin Yoon, Nanxuan Zhao, Julien Philip, Anna Fr{\"u}hst{\"u}ck, Wenbin Li, Christian Richardt, and Tuanfeng Wang.
\newblock Intrinsic{D}iffusion: Joint intrinsic layers from latent diffusion models.
\newblock In \emph{ACM SIGGRAPH Conference Papers}, pages 1--11, 2024.

\bibitem[Ma et~al.(2023)Ma, Xu, Zhang, Zhou, and Wu]{ma2023opensvbrdf}
Xiaohe Ma, Xianmin Xu, Leyao Zhang, Kun Zhou, and Hongzhi Wu.
\newblock Open{SVBRDF}: A database of measured spatially-varying reflectance.
\newblock \emph{ACM TOG}, 42\penalty0 (6):\penalty0 1--14, 2023.

\bibitem[Marschner et~al.(2003)Marschner, Jensen, Cammarano, Worley, and Hanrahan]{marschner2003light}
Stephen~R Marschner, Henrik~Wann Jensen, Mike Cammarano, Steve Worley, and Pat Hanrahan.
\newblock Light scattering from human hair fibers.
\newblock \emph{ACM TOG}, 22\penalty0 (3):\penalty0 780--791, 2003.

\bibitem[Montazeri et~al.(2020)Montazeri, Gammelmark, Zhao, and Jensen]{montazeri2020practical}
Zahra Montazeri, S{\o}ren~B Gammelmark, Shuang Zhao, and Henrik~Wann Jensen.
\newblock A practical ply-based appearance model of woven fabrics.
\newblock \emph{ACM TOG}, 39\penalty0 (6):\penalty0 1--13, 2020.

\bibitem[Oren and Nayar(1994)]{oren1994generalization}
Michael Oren and Shree~K Nayar.
\newblock Generalization of {L}ambert's reflectance model.
\newblock In \emph{ACM SIGGRAPH Conference Papers}, pages 239--246, 1994.

\bibitem[Pharr and Fernando(2005)]{pharr2005gpu}
Matt Pharr and Randima Fernando.
\newblock \emph{GPU Gems 2: Programming techniques for high-performance graphics and general-purpose computation (GPU Gems)}.
\newblock Addison-Wesley Professional, 2005.

\bibitem[Pharr et~al.(2023)Pharr, Jakob, and Humphreys]{pharr2023physically}
Matt Pharr, Wenzel Jakob, and Greg Humphreys.
\newblock \emph{Physically based rendering: From theory to implementation}.
\newblock MIT Press, 2023.

\bibitem[Phongthawee et~al.(2024)Phongthawee, Chinchuthakun, Sinsunthithet, Jampani, Raj, Khungurn, and Suwajanakorn]{phongthawee2024diffusionlight}
Pakkapon Phongthawee, Worameth Chinchuthakun, Nontaphat Sinsunthithet, Varun Jampani, Amit Raj, Pramook Khungurn, and Supasorn Suwajanakorn.
\newblock Diffusion{L}ight: Light probes for free by painting a chrome ball.
\newblock In \emph{CVPR}, pages 98--108, 2024.

\bibitem[Ramamoorthi and Hanrahan(2001)]{ramamoorthi2001efficient}
Ravi Ramamoorthi and Pat Hanrahan.
\newblock An efficient representation for irradiance environment maps.
\newblock In \emph{ACM SIGGRAPH Conference Papers}, pages 497--500, 2001.

\bibitem[Ramesh et~al.(2022)Ramesh, Dhariwal, Nichol, Chu, and Chen]{ramesh2022hierarchical}
Aditya Ramesh, Prafulla Dhariwal, Alex Nichol, Casey Chu, and Mark Chen.
\newblock Hierarchical text-conditional image generation with clip latents.
\newblock \emph{arXiv preprint arXiv:2204.06125}, 1\penalty0 (2):\penalty0 3, 2022.

\bibitem[Rodr{\'\i}guez-Pardo et~al.(2025)Rodr{\'\i}guez-Pardo, Pascual-Hernandez, Rodriguez-Vazquez, Lopez-Moreno, and Garces]{rodriguez2025single}
Carlos Rodr{\'\i}guez-Pardo, David Pascual-Hernandez, Javier Rodriguez-Vazquez, Jorge Lopez-Moreno, and Elena Garces.
\newblock Single-image reflectance and transmittance estimation from any flatbed scanner.
\newblock \emph{Computers \& Graphics}, page 104186, 2025.

\bibitem[Rombach et~al.(2022)Rombach, Blattmann, Lorenz, Esser, and Ommer]{rombach2022high}
Robin Rombach, Andreas Blattmann, Dominik Lorenz, Patrick Esser, and Bj{\"o}rn Ommer.
\newblock High-resolution image synthesis with latent diffusion models.
\newblock In \emph{CVPR}, pages 10684--10695, 2022.

\bibitem[Schlick(1994)]{schlick1994inexpensive}
Christophe Schlick.
\newblock An inexpensive {BRDF} model for physically-based rendering.
\newblock \emph{Computer Graphics Forum}, 13\penalty0 (3):\penalty0 233--246, 1994.

\bibitem[Sharma et~al.(2024)Sharma, Jampani, Li, Jia, Lagun, Durand, Freeman, and Matthews]{sharma2024alchemist}
Prafull Sharma, Varun Jampani, Yuanzhen Li, Xuhui Jia, Dmitry Lagun, Fredo Durand, Bill Freeman, and Mark Matthews.
\newblock Alchemist: Parametric control of material properties with diffusion models.
\newblock In \emph{CVPR}, pages 24130--24141, 2024.

\bibitem[Siddiqui et~al.(2025)Siddiqui, Monnier, Kokkinos, Kariya, Kleiman, Garreau, Gafni, Neverova, Vedaldi, Shapovalov, et~al.]{siddiqui2025meta}
Yawar Siddiqui, Tom Monnier, Filippos Kokkinos, Mahendra Kariya, Yanir Kleiman, Emilien Garreau, Oran Gafni, Natalia Neverova, Andrea Vedaldi, Roman Shapovalov, et~al.
\newblock Meta 3d assetgen: Text-to-mesh generation with high-quality geometry, texture, and pbr materials.
\newblock In \emph{NeurIPS}, pages 9532--9564, 2025.

\bibitem[UE5()]{ue5}
UE5.
\newblock Unreal {E}ngine 5.5: Physically based materials.

\bibitem[Veach(1998)]{veach1998robust}
Eric Veach.
\newblock \emph{Robust Monte Carlo methods for light transport simulation}.
\newblock Stanford University, 1998.

\bibitem[Walter et~al.(2007)Walter, Marschner, Li, and Torrance]{walter2007microfacet}
Bruce Walter, Stephen~R Marschner, Hongsong Li, and Kenneth~E Torrance.
\newblock Microfacet models for refraction through rough surfaces.
\newblock \emph{Rendering techniques}, 2007:\penalty0 18th, 2007.

\bibitem[Wang et~al.(2022)Wang, Jin, Ha{\v{s}}an, and Yan]{wang2022spongecake}
Beibei Wang, Wenhua Jin, Milo{\v{s}} Ha{\v{s}}an, and Ling-Qi Yan.
\newblock Spongecake: A layered microflake surface appearance model.
\newblock \emph{ACM TOG}, 42\penalty0 (1):\penalty0 1--16, 2022.

\bibitem[Wang et~al.(2024{\natexlab{a}})Wang, Zhang, Gao, Kang, and Zhang]{wang2024nfplight}
Li Wang, Lianghao Zhang, Fangzhou Gao, Yuzhen Kang, and Jiawan Zhang.
\newblock {NFPLight}: Deep {SVBRDF} estimation via the combination of near and far field point lighting.
\newblock \emph{ACM TOG}, 43\penalty0 (6):\penalty0 1--11, 2024{\natexlab{a}}.

\bibitem[Wang et~al.(2025)Wang, Tran, Cui, TG, Chandraker, and Frisvad]{wang2025materialist}
Lezhong Wang, Duc~Minh Tran, Ruiqi Cui, Thomson TG, Manmohan Chandraker, and Jeppe~Revall Frisvad.
\newblock Materialist: Physically based editing using single-image inverse rendering.
\newblock \emph{arXiv preprint arXiv:2501.03717}, 2025.

\bibitem[Wang et~al.(2024{\natexlab{b}})Wang, Xu, Ma, Wang, and Dai]{wang2024boosting}
Yitong Wang, Xudong Xu, Li Ma, Haoran Wang, and Bo Dai.
\newblock Boosting 3d object generation through {PBR} materials.
\newblock In \emph{SIGGRAPH Asia Conference Papers}, pages 1--11, 2024{\natexlab{b}}.

\bibitem[Yang et~al.(2024{\natexlab{a}})Yang, Kang, Huang, Xu, Feng, and Zhao]{yang2024depth}
Lihe Yang, Bingyi Kang, Zilong Huang, Xiaogang Xu, Jiashi Feng, and Hengshuang Zhao.
\newblock Depth anything: Unleashing the power of large-scale unlabeled data.
\newblock In \emph{CVPR}, pages 10371--10381, 2024{\natexlab{a}}.

\bibitem[Yang et~al.(2024{\natexlab{b}})Yang, Kang, Huang, Zhao, Xu, Feng, and Zhao]{yang2025depth}
Lihe Yang, Bingyi Kang, Zilong Huang, Zhen Zhao, Xiaogang Xu, Jiashi Feng, and Hengshuang Zhao.
\newblock Depth anything v2.
\newblock In \emph{NeurIPS}, pages 21875--21911, 2024{\natexlab{b}}.

\bibitem[Ye et~al.(2024)Ye, Qiu, Gu, Zuo, Wu, Dong, Bo, Xiu, and Han]{ye2024stablenormal}
Chongjie Ye, Lingteng Qiu, Xiaodong Gu, Qi Zuo, Yushuang Wu, Zilong Dong, Liefeng Bo, Yuliang Xiu, and Xiaoguang Han.
\newblock Stablenormal: Reducing diffusion variance for stable and sharp normal.
\newblock \emph{ACM TOG}, 43\penalty0 (6):\penalty0 1--18, 2024.

\bibitem[Zeng et~al.(2024)Zeng, Deschaintre, Georgiev, Hold-Geoffroy, Hu, Luan, Yan, and Ha\v{s}an]{zeng2024rgb}
Zheng Zeng, Valentin Deschaintre, Iliyan Georgiev, Yannick Hold-Geoffroy, Yiwei Hu, Fujun Luan, Ling-Qi Yan, and Milo\v{s} Ha\v{s}an.
\newblock {RGBX}: Image decomposition and synthesis using material- and lighting-aware diffusion models.
\newblock In \emph{ACM SIGGRAPH Conference Papers}, 2024.

\bibitem[Zhang et~al.(2023)Zhang, Rao, and Agrawala]{zhang2023adding}
Lvmin Zhang, Anyi Rao, and Maneesh Agrawala.
\newblock Adding conditional control to text-to-image diffusion models.
\newblock In \emph{CVPR}, pages 3836--3847, 2023.

\bibitem[Zhang et~al.(2025)Zhang, Rao, and Agrawala]{zhang2025scaling}
Lvmin Zhang, Anyi Rao, and Maneesh Agrawala.
\newblock Scaling in-the-wild training for diffusion-based illumination harmonization and editing by imposing consistent light transport.
\newblock In \emph{ICLR}, 2025.

\bibitem[Zhang et~al.(2018)Zhang, Isola, Efros, Shechtman, and Wang]{zhang2018unreasonable}
Richard Zhang, Phillip Isola, Alexei~A Efros, Eli Shechtman, and Oliver Wang.
\newblock The unreasonable effectiveness of deep features as a perceptual metric.
\newblock In \emph{CVPR}, pages 586--595, 2018.

\bibitem[Zhang et~al.(2024)Zhang, Liu, Xie, Yang, Liu, Yang, Zhang, Kou, Lin, Wang, et~al.]{zhang2024dreammat}
Yuqing Zhang, Yuan Liu, Zhiyu Xie, Lei Yang, Zhongyuan Liu, Mengzhou Yang, Runze Zhang, Qilong Kou, Cheng Lin, Wenping Wang, et~al.
\newblock Dreammat: High-quality {PBR} material generation with geometry-and light-aware diffusion models.
\newblock \emph{ACM TOG}, 43\penalty0 (4):\penalty0 1--18, 2024.

\bibitem[Zhu et~al.(2022)Zhu, Luan, Huo, Lin, Zhong, Xi, Wang, Bao, Zheng, and Tang]{zhu2022learning}
Jingsen Zhu, Fujun Luan, Yuchi Huo, Zihao Lin, Zhihua Zhong, Dianbing Xi, Rui Wang, Hujun Bao, Jiaxiang Zheng, and Rui Tang.
\newblock Learning-based inverse rendering of complex indoor scenes with differentiable {M}onte {C}arlo raytracing.
\newblock In \emph{SIGGRAPH Asia Conference Papers}, pages 1--8, 2022.

\bibitem[Zhu et~al.(2024)Zhu, Qiu, Gu, Zhao, Xu, He, Li, Han, Yao, Cao, et~al.]{zhu2024mcmat}
Shenhao Zhu, Lingteng Qiu, Xiaodong Gu, Zhengyi Zhao, Chao Xu, Yuxiao He, Zhe Li, Xiaoguang Han, Yao Yao, Xun Cao, et~al.
\newblock {MCMat}: Multiview-consistent and physically accurate pbr material generation.
\newblock \emph{arXiv preprint arXiv:2412.14148}, 2024.

\bibitem[Zhuang et~al.(2021)Zhuang, Shen, Wang, and Liu]{zhuang2021real}
Tao Zhuang, Pengfei Shen, Beibei Wang, and Ligang Liu.
\newblock Real-time denoising using {BRDF} pre-integration factorization.
\newblock \emph{Computer Graphics Forum}, 40\penalty0 (7):\penalty0 173--180, 2021.

\end{thebibliography}
}


\end{document}